\documentclass[aps,prx,showpacs,floatfix,twocolumn,superscriptaddress,footinbib]{revtex4-2}

\usepackage{bm}
\usepackage[version=4]{mhchem}
\usepackage{glossaries}
\usepackage{hyperref}
\usepackage{orcidlink}

\hypersetup{
    colorlinks=true,
    linkcolor=blue,
    filecolor=magenta,
    urlcolor=red,
    citecolor=blue,
}

\def\Q{\ensuremath{\bm{Q}}}
\newcommand{\angstrom}{\mbox{\normalfont\AA}}

\newacronym{RP}{RP}{Ruddlesden-Popper}
\newacronym{RIXS}{RIXS}{resonant inelastic x-ray scattering}
\newacronym{XAS}{XAS}{x-ray absorption spectroscopy}
\newacronym{EELS}{EELS}{electron energy-loss spectroscopy}
\newacronym{FWHM}{FWHM}{full-width at half-maximum}
\newacronym{HWHM}{HWHM}{half-width at half-maximum}
\newacronym{AIM}{AIM}{Anderson impurity model}
\newacronym{ED}{ED}{exact diagonalization}
\newacronym{AFM}{AFM}{antiferromagnetic}
\newacronym{FM}{FM}{ferromagnetic}
\newacronym{DFT}{DFT}{density functional theory}
\newacronym{DMFT}{DMFT}{dynamical mean-field theory}
\newacronym{2D}{2D}{two-dimensional}
\newacronym{ZSA}{ZSA}{Zaanen-Sawatzky-Allen}
\newacronym{DHO}{DHO}{damped harmonic oscillator}
\newacronym{EPC}{EPC}{electron-phonon coupling}
\newacronym{ARPES}{ARPES}{angle-resolved photoemission spectroscopy}
\newacronym{RPA}{RPA}{random phase approximation}

\begin{document}

\title{Layer Architecture Shapes Electronic, Magnetic, and Lattice Interactions in Ruddlesden-Popper Nickelates}

\author{W.~He\,\orcidlink{0000-0003-3522-3899}}
\affiliation{Condensed Matter Physics and Materials Science Department, Brookhaven National Laboratory, Upton, New York 11973, USA}
\affiliation{Stanford Institute for Materials and Energy Sciences, SLAC National Accelerator Laboratory, Menlo Park, California 94025, USA}
\author{X.~Guo\,\orcidlink{0000-0003-1088-6039}}
\author{X.~Luo\,\orcidlink{0000-0002-5471-053X}}
\affiliation{Condensed Matter Physics and Materials Science Department, Brookhaven National Laboratory, Upton, New York 11973, USA}

\author{J.~Thomas\,\orcidlink{0000-0003-4818-6660}}
\affiliation{Department of Physics and Astronomy, The University of Tennessee, Knoxville, Tennessee 37996, USA}
\affiliation{Institute for Advanced Materials and Manufacturing, University of Tennessee, Knoxville, Tennessee 37996, USA\looseness=-1}

\author{J.~Sears\,\orcidlink{0000-0001-6524-8953}}
\affiliation{Condensed Matter Physics and Materials Science Department, Brookhaven National Laboratory, Upton, New York 11973, USA}

\author{Sophia~F.~R.~TenHuisen\,\orcidlink{0000-0002-4379-7035}}
\author{Ziqiang~Guan\,\orcidlink{0000-0002-1097-6766}}
\affiliation{Department of Physics, Harvard University, Cambridge, Massachusetts 02138, USA}
\author{Xinglong~Chen\,\orcidlink{0000-0003-2283-7293}}
\affiliation{Materials Science Division, Argonne National Laboratory, Lemont, Illinois 60439, USA}

\author{D.~A.~Dahlbom\,\orcidlink{0000-0002-0221-5086}}
\affiliation{Neutron Scattering Division, Oak Ridge National Laboratory, Oak Ridge, Tennessee, USA}

\author{B.~Zager\,\orcidlink{0000-0001-7436-4040}}
\author{J.~Pelliciari\,\orcidlink{0000-0003-1508-7746}}
\affiliation{National Synchrotron Light Source II, Brookhaven National Laboratory, Upton, New York 11973, USA}

\author{Yi-Feng~Zhao}
\affiliation{Department of Physics and Astronomy, Arizona State University, Tempe, AZ 85218, USA}
\author{H.~LaBollita\,\orcidlink{0000-0002-6699-8577}}
\affiliation{Center for Computational Quantum Physics, Flatiron Institute, New York, NY 10010, USA}

\author{Hong~Zheng\,\orcidlink{0000-0002-6342-7277}}
\affiliation{Materials Science Division, Argonne National Laboratory, Lemont, Illinois 60439, USA}

\author{M.~K.~Lajer\,\orcidlink{0000-0002-1168-8598}}
\affiliation{Condensed Matter Physics and Materials Science Department, Brookhaven National Laboratory, Upton, New York 11973, USA}

\author{J.~F.~Mitchell\,\orcidlink{0000-0002-8416-6424}}
\affiliation{Materials Science Division, Argonne National Laboratory, Lemont, Illinois 60439, USA}

\author{V.~Bisogni\,\orcidlink{0000-0002-7399-9930}}
\affiliation{National Synchrotron Light Source II, Brookhaven National Laboratory, Upton, New York 11973, USA}

\author{A.~S.~Botana}
\affiliation{Department of Physics and Astronomy, Arizona State University, Tempe, AZ 85218, USA}

\author{M.~Mitrano\,\orcidlink{0000-0002-0102-0391}}
\affiliation{Department of Physics, Harvard University, Cambridge, Massachusetts 02138, USA}

\author{S.~Johnston\,\orcidlink{0000-0002-2343-0113}}\email[]{sjohn145@utk.edu}
\affiliation{Department of Physics and Astronomy, The University of Tennessee, Knoxville, Tennessee 37996, USA}
\affiliation{Institute for Advanced Materials and Manufacturing, University of Tennessee, Knoxville, Tennessee 37996, USA\looseness=-1}

\author{M.~P.~M.~Dean\,\orcidlink{0000-0001-5139-3543}}\email[]{mdean@bnl.gov}
\affiliation{Condensed Matter Physics and Materials Science Department, Brookhaven National Laboratory, Upton, New York 11973, USA}
\affiliation{Department of Physics and Astronomy, The University of Tennessee, Knoxville, Tennessee 37996, USA}

\date{\today}

\begin{abstract}

The discovery of superconductivity in Ruddlesden–Popper nickelates has raised a central question: how does layer architecture shape the electronic, magnetic, and lattice interactions relevant to pairing? Here, we report a detailed comparative study of the two polymorphs of \ce{La3Ni2O7}---the alternating monolayer-trilayer (LNO-1313) and bilayer (LNO-2222) structures---and the related trilayer compound \ce{La4Ni3O10}, using both Ni $L_3$- and O $K$-edge \gls*{RIXS}. We find that LNO-1313 and \ce{La4Ni3O10} share strikingly similar electronic, magnetic, and lattice excitations, whereas bilayer LNO-2222 exhibits distinct features. Compared to LNO-2222, LNO-1313 and \ce{La4Ni3O10} have weaker orbital polarization, enhanced $3d^8\underline{L}$ character, a reduced out-of-plane magnetic-exchange scale, and stronger \gls*{EPC}. Within an effective local-moment framework, an entangled-dimer scenario provides a natural description of the spin excitations generated by strong antiferromagnetic interlayer coupling. Its advantage over conventional spin-wave theory is clearest in bilayer LNO-2222, where the interlayer coupling dominates the intralayer interactions. These findings provide critical experimental constraints for future theoretical models for the low-energy physics relevant to superconductivity in these layered nickelates.

\end{abstract}

\maketitle
\glsresetall  

\section{Introduction}

The discovery of superconductivity in square-planar nickelates $R_{n+1}$Ni$_n$O$_{2n+2}$ ($R=$ rare-earth ion) established nickelates as a new family of unconventional superconductors \cite{Li2019Superconductivity, Pan2026Superconducting}. In parallel, perovskite-type \gls*{RP} nickelates $R_{n+1}$Ni$_n$O$_{3n+1}$ have emerged as a promising platform for exploring high-temperature superconductivity \cite{Sun2023Signatures, Zhang2024Hightemperature, Zhu2024Superconductivity, Zhang2025Bulk, Shi2025Pressure, Ko2025Signatures, Zhou2025Ambientpressure, Liu2025Superconductivity}. In particular, bilayer \ce{La3Ni2O7} \cite{Sun2023Signatures, Zhang2024Hightemperature} and trilayer \ce{La4Ni3O10} \cite{Zhu2024Superconductivity} are the two most studied systems and realize maximum superconducting transition temperatures $T_\mathrm{c}$ of up to 80~K and 30~K, respectively, under pressure or epitaxial strain. This reduction of $T_\mathrm{c}$ from bilayer to trilayer---opposite to the trend in cuprates---highlights the need to understand how different structural layering affects the low-energy electronic structure and the underlying pairing correlations. Notably, in addition to the common bilayer structure found in \ce{La3Ni2O7} (referred to as LNO-2222 hereafter), \ce{La3Ni2O7} can also crystallize in a ``1313'' polymorph consisting of alternating monolayer-trilayer blocks (referred to as LNO-1313 hereafter) \cite{Chen2024Polymorphism, Puphal2024Unconventional, Wang2024Longrange}. Although signs of filamentary superconductivity have been reported in this new polymorph under pressure \cite{Puphal2024Unconventional},
it remains unclear whether LNO-1313 can also host bulk superconductivity \cite{Puphal2024Unconventional, Abadi2025Electronic}. Comparing this alternating-layer structure to its uniform trilayer cousin \ce{La4Ni3O10} offers a unique opportunity to examine the impact of an inserted single \ce{La2NiO4} layer.

\Gls*{RIXS} is a versatile technique for simultaneously probing electronic, magnetic, and lattice excitations in strongly correlated systems with elemental and orbital sensitivity \cite{Mitrano2024exploring, Hepting2021Soft}. In \gls*{RP} nickelates, Ni $L_3$-edge \gls*{RIXS} has been utilized to probe electronic and magnetic excitations in bilayer LNO-2222 \cite{Chen2024Electronic, Zhong2026Spin, Zhong2026Doping, Zhang2026Interlayer, Li2026Interlayer, Chen2026Orbital} and trilayer \ce{La4Ni3O10} \cite{Fabbris2023Resonant} and \ce{Nd4Ni3O10} \cite{TenHuisen2025Magnetic}, offering key insights into their electronic structure and magnetic-excitation dispersions. By contrast, the alternating-layer LNO-1313 has remained unexplored by \gls*{RIXS}, leaving a critical gap in understanding how its electronic and magnetic excitations relate to those of bilayer LNO-2222 and trilayer \ce{La4Ni3O10}.

Here, we present a comprehensive \gls*{RIXS} study of the two \ce{La3Ni2O7} polymorphs and \ce{La4Ni3O10} single crystals at ambient pressure, performed at both Ni $L$- and O $K$-edges, with a focus on their low-energy electronic, magnetic, and lattice excitations. Spin fluctuations and \gls*{EPC}, two primary candidates for superconducting pairing ``glue(s)'' in unconventional superconductors \cite{Scalapino2012common, Puphal2026superconductivity}, can both be evaluated in this work through their respective magnetic and phonon excitations in the \gls*{RIXS} spectra. Our results reveal striking similarities between LNO-1313 and \ce{La4Ni3O10} across all three degrees of freedom, in sharp contrast to the behavior of the bilayer variant LNO-2222. These findings indicate that the trilayer motifs strongly influence the low-energy excitation spectrum in LNO-1313 and provide valuable constraints for theoretical models of the electronic and magnetic structures of these materials. In particular, strong Hund's coupling is crucial for promoting $3d^8\underline{L}$ character in the ground state of all three compounds by driving the self-doped holes onto the ligand sites. Meanwhile, the active $3d_{3z^2-r^2}$ orbitals give rise to strong interlayer coupling, which we argue promotes the formation of interlayer dimers with reduced ordered moments and enhanced spin fluctuations.

\section{Methods}
Single crystals of \ce{La4Ni3O10}, LNO-1313, and LNO-2222 were grown using a high-pressure floating-zone technique in an oxygen atmosphere \cite{Zhang2020High, Chen2024Polymorphism}. X-ray diffraction measurements (see Supplemental Figs.~S1 and S2) confirmed the phase purity of all three samples. These materials crystallize with slightly different symmetries: monoclinic $P2_1/a$ (space group No.~14) for \ce{La4Ni3O10} \cite{Zhang2020High}, orthorhombic $Cmmm$ (space group No.~65) for LNO-1313 \cite{Chen2024Polymorphism}, and orthorhombic $Amam$ (space group No.~63) for LNO-2222 \footnote{The source of our samples found a space group of $Amam$ \cite{Chen2024Polymorphism}, which is the result we quote. A recent preprint suggests a lower symmetry of $Am2m$ \cite{Misawa2026polar}.}. For simplicity, we adopted a pseudo-tetragonal unit-cell notation throughout this manuscript. The lattice parameters following this convention are $a = b = 3.84$~\angstrom{} and $c = 27.96$~\angstrom{} for \ce{La4Ni3O10}, $a = b = 3.85$~\angstrom{} and $c = 20.3$~\angstrom{} for LNO-1313, and $a = b = 3.833$~\angstrom{} and $c = 20.45$~\angstrom{} for LNO-2222. We then index the momentum transfer $\Q{}=(H,K,L)$ in reciprocal lattice units (r.l.u.) for each sample. To compare the out-of-plane periodicities of the different compounds, we also define the effective coordinate $L^*=Ld/c$, where $d\approx 3.9$~\AA{} is the separation between neighboring \ce{NiO2} layers in the structure. We also note that the different stoichiometries lead to different nominal electron counts of $d^{7.5}$ for LNO-2222 and LNO-1313 compared to $d^{7.33}$ in \ce{La4Ni3O10}.

High-energy-resolution \gls*{RIXS} measurements were performed at the SIX 2-ID beamline of the National Synchrotron Light Source II \cite{Dvorak2016Towards}. Samples were cleaved with the $c$-axis surface normal and pre-aligned using a laboratory single-crystal x-ray diffractometer. For all three materials, two orthogonal structural twin domains have been observed with similar populations, indicating equivalence between $(HHL)$ and $(H\bar{H}L)$ scattering planes. Incident-energy-dependent Ni $L_3$-edge \gls*{RIXS} measurements were taken with both linear horizontal ($\pi$) and vertical ($\sigma$) polarizations in the $(H0L)$ scattering plane with an incident angle of $\theta = 10^\circ$ at $T=80$~K. The in-plane momentum-dependent \gls*{RIXS} data were collected with $\pi$-polarized incident photons at the resonant energy for magnetic excitations of $\sim 852.9$~eV at $T=80$~K. The out-of-plane momentum-dependent \gls*{RIXS} spectra were measured at $T=22$~K. The spectrometer was operated with an energy resolution of $32$~meV \gls*{FWHM} for the in-plane momentum dependence and a slightly relaxed resolution of $38$~meV for the incident-energy and out-of-plane momentum dependences. The O $K$-edge \gls*{RIXS} measurements were performed with both $\pi$- and $\sigma$-polarized photons at an incident angle of $\theta = 15^\circ$ and at $T=22$~K, using an energy resolution of $27$~meV. The scattering angle was kept at $2\Theta = 150^\circ$ except during the out-of-plane momentum dependence measurements. An angle-dependent correction factor was applied to all \gls*{RIXS} spectra to account for self-absorption effects \cite{Miao2017High}. Where appropriate, spectra are normalized to counting time before comparison.

\section{Electronic structure}

\begin{figure*}
\includegraphics{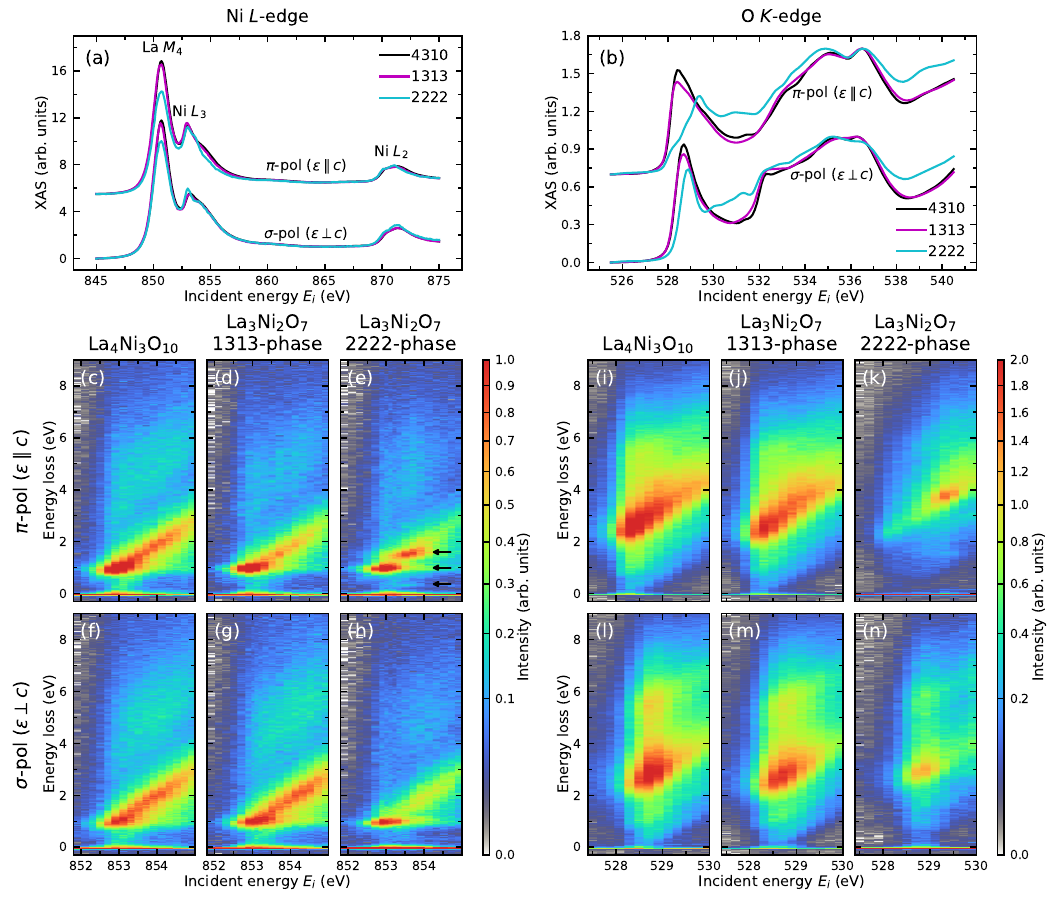}
\caption{Electronic structure of \ce{La4Ni3O10} and the two polymorphs of \ce{La3Ni2O7}. (a),(b) \gls*{XAS} spectra measured in total fluorescence yield mode for (a) the Ni $L$-edge and (b) the O $K$-edge, respectively. (c)--(n), \gls*{RIXS} intensity maps as a function of incident x-ray energy across (c)--(h) the Ni $L_3$ and (i)--(n) the O $K$ resonances, respectively. The Ni $L$-edge data were taken at 80~K with an incident angle of $\theta = 10^\circ$ and scattering angle of $2\Theta = 150^\circ$ in the ($H0L$) scattering plane, while the O $K$-edge data were taken at 22~K with an incident angle of $\theta = 15^\circ$ and scattering angle of $2\Theta = 150^\circ$ in the ($HHL$) scattering plane. Therefore, the polarization of $\sigma$($\pi$)-polarized incident x-rays is perpendicular (approximately parallel) to the sample $c$-axis. The \gls*{XAS} $L$-edge data are normalized to the post-$L_3$-edge value in each spectrum. Due to the very extended O $K$-edge spectrum, these spectra are normalized to their maximum value. $\epsilon\parallel c$ spectra are vertically shifted for clarity. The arrows in panel (e) indicate the three $dd$-excitation peaks in LNO-2222.
}
\label{fig:energy_map}
\end{figure*}

We start by assessing the electronic structure of our chosen family of nickelates. We present polarization-dependent measurements of LNO-1313 at the Ni $L$ and O $K$ edges, compared with corresponding spectra from \ce{La4Ni3O10} and LNO-2222, as displayed in Fig.~\ref{fig:energy_map}. The Ni $L_3$-edge in Fig.~\ref{fig:energy_map}(a) overlaps with the strong La $M_4$ resonance near 850~eV, while the Ni $L_2$-edge is spectrally well-separated from other absorption edges. At both Ni edges, two peaks separated by approximately $1$~eV are evident, most clearly in LNO-2222. Similar to other types of perovskite nickelates, the lower (higher) energy \gls*{XAS} peaks correspond to configurations with smaller (larger) oxygen-ligand contribution \cite{Bisogni2016Groundstate, Fabbris2017Doping}. The O $K$-edge spectra can be separated into a ``pre-peak'' region from 527 to 531~eV that involves Ni $3d$--O $2p$ hybridized holes and a ``main edge'' corresponding to states with leading oxygen character. The \gls*{XAS} data for \ce{La4Ni3O10} and LNO-2222 are consistent with prior reports \cite{Chen2024Electronic, Ren2025Resolving, Zhong2026Spin, Zhang2017Large, Fabbris2023Resonant}.

A striking similarity between the spectra for \ce{La4Ni3O10} and LNO-1313 is immediately apparent. Another notable observation is the relatively small difference in intensity between the two incident x-ray polarizations (see the direct comparison of polarization-dependent \gls*{XAS} in Supplemental Fig.~S3), especially for \ce{La4Ni3O10} and LNO-1313, in sharp contrast to square-planar nickelates \cite{Zhang2017Large, Rossi2021Orbital} or cuprates \cite{Chen1992Outofplane}. Such weak orbital polarization indicates that holes are distributed across $3d_{x^2-y^2}$ and $3d_{3z^2-r^2}$ orbitals with comparable occupancy.

The substantial intensity of configurations with strong ligand character, such as the features at 854 and 872~eV, reflects strong hybridization between the Ni $3d$ and O $2p$ states in these materials [Fig.~\ref{fig:energy_map}(a)]. This is further evidenced by the strong prepeak intensity around $528.7$~eV at the O $K$-edge shown in Fig.~\ref{fig:energy_map}(b). Previous atomically resolved \gls*{EELS} on LNO-2222 \cite{Dong2024Visualization} has revealed two major ligand-hole destinations, i.e., the inner apical O and planar O sites, with negligible contributions from the outer apical oxygens. Therefore, we infer that the extended spectral weight in the prepeak region in the $\pi$-polarization channel arises mostly from hybridized inner apical O $2p_z$ and Ni $3d_{3z^2-r^2}$ orbitals, while the narrower peak in the $\sigma$-polarization channel arises from hybridized planar O $2p_{x,y}$ and Ni $3d_{x^2-y^2}$ orbitals, consistent with the conclusions in Ref.~\cite{Ren2025Resolving}. The wider oxygen bandwidth along the out-of-plane direction in all three compounds could be caused by the formation of strongly hybridized molecular orbitals via interlayer hopping along inner apical O $2p_z$. We also note that the leading edges of the $\pi$-polarized data always appear slightly lower than those of the corresponding $\sigma$-polarized spectra at both the Ni and O edges for all three nickelates (by approximately $0.2$~eV, see Supplemental Fig.~S3), indicating subtle energy differences between the $3d_{3z^2-r^2}$ and $3d_{x^2-y^2}$ orbitals.

Incident-energy-dependent \gls*{RIXS} maps provide higher-dimensional information than \gls*{XAS} spectra, enabling a more comprehensive interrogation of materials' electronic structure. Figures~\ref{fig:energy_map}(c)--\ref{fig:energy_map}(h) compare the Ni $L_3$-edge \gls*{RIXS} maps of LNO-1313 with those of the other nickelates. The data for \ce{La4Ni3O10} and LNO-2222 are consistent with previous reports on bulk materials \cite{Fabbris2023Resonant, Chen2024Electronic}, but were acquired with improved energy resolution and/or an extended energy-loss range. O $K$-edge \gls*{RIXS} maps [Figs.~\ref{fig:energy_map}(i)--\ref{fig:energy_map}(n)] enable direct characterization of the hybridized states between Ni $3d$ and O $2p$ orbitals \cite{Shen2022Role}. All three materials exhibit three distinct types of features spanning the sub-eV to several-eV energy scale: (1) relatively sharp $dd$-type excitations at fixed energy loss below $2$~eV; (2) broader charge-transfer-related intensity in the $\sim2.5$--$8$~eV window; and (3) fluorescence lines extending along the diagonal direction. Low-energy excitations are also present and will be discussed later, as they can only be seen in plots that zoom in on the low-energy region of the spectra.

Ni $L_3$-edge \gls*{RIXS} offers enhanced cross-sections for localized $dd$-excitations. Careful examination of Fig.~\ref{fig:energy_map}(e) reveals three peaks in LNO-2222 at energy losses of approximately $0.4$, $1.0$, and $1.6$~eV. The $1.0$~eV peak is stronger at the $3d^8$ resonance with minimal polarization dependence, consistent with the previous assignment of crystal-field excitations between Ni $3d$ $t_{\mathrm{2g}}$ and $e_{\mathrm{g}}$ orbitals \cite{Chen2024Electronic}. In contrast, the $1.6$ and $0.4$~eV excitations resonate with the $3d^8\underline{L}$ manifolds and manifest pronounced dichroism, with stronger signals observed in the $\pi$-polarized data. The $1.6$~eV peak has been assigned to $dd$-type transitions involving the full Ni $3d$ manifold, whereas the $0.4$~eV peak has been attributed to transitions between the Ni $3d_{3z^2-r^2}$ and $3d_{x^2-y^2}$ orbitals \cite{Chen2024Electronic}. Due to the above-mentioned resonance and polarization behaviors, we propose that hybridization between the inner apical O $2p_z$ and Ni $3d_{3z^2-r^2}$ orbitals may play an important role in shaping the intensities of these excitations. We observe significant changes in these $dd$-type excitations in \ce{La4Ni3O10} and LNO-1313. The $1.6$~eV peak exhibits substantial broadening and merges with the fluorescence line, whereas the $0.4$~eV peak diminishes and is replaced by accumulated spectral weight near the $3d^8$ resonance energy, rather than the higher-energy $3d^8\underline{L}$ resonance. The $1.0$~eV peak also broadens slightly. Overall, the markedly altered peak profiles with greatly reduced orbital polarization closely resemble those observed in \ce{La_{2-x}Sr_{x}NiO4} \cite{Fabbris2017Doping} and $R$\ce{NiO3} \cite{Bisogni2016Groundstate, Fabbris2016Orbital}, where the ground state is predominantly $3d^8\underline{L}$ in character. This similarity suggests enhanced $3d^8\underline{L}$ character in \ce{La4Ni3O10} and LNO-1313, which might be related to the increased effective doping level from bilayer to trilayer. The emergence of the broad continuum around $0.4$~eV is also reminiscent of the metallic phase of \ce{NdNiO3} \cite{Bisogni2016Groundstate}, suggesting enhanced metallic behavior relative to LNO-2222.

Charge-transfer excitations are more effectively probed by O $K$-edge \gls*{RIXS} because this resonance couples directly to O-character transitions. In contrast, the $dd$ excitations at the O $K$-edge rely on Ni--O hybridization and are therefore less intense than the charge-transfer excitations. This distinction in intensity helps us identify the charge-transfer excitations as the strongest features in panels (i)--(n) of Fig.~\ref{fig:energy_map}. The energy required for a charge-transfer excitation, in this case between $d^8$ and $d^9\underline{L}$ states (where $\underline{L}$ is a ligand hole), is somewhat similar to the energy required for $d^9 \rightarrow d^{10} \underline{L}$ charge-transfer excitations in cuprates and low-valence nickelates \cite{Shen2022Role, Shen2023Electronic}. This supports an active role for oxygen states in all these materials. Besides the strongest peak around $2.5$~eV detected at the O $K$-edge, the charge-transfer-related feature extends to the $\sim4$--$6$~eV energy-loss region at the main O and Ni resonances. Such a feature is well captured in recent \gls*{RIXS} simulations at the Ni edge \cite{Norman2023Orbital} for \ce{La4Ni3O10}. The $2.5$~eV peak is less discernible at the Ni edge in both our experimental data and previous simulations \cite{Norman2023Orbital}, likely because it is obscured by the stronger nearby $dd$ and fluorescence features, together with the intrinsically weaker cross-section for charge-transfer excitations at the transition-metal edge. Future calculations at the O $K$-edge could help further confirm the nature of this $2.5$~eV peak.

Lastly, the fluorescence features, observed at fixed emitted photon energies, originate from excitations into a continuum formed by more extended states, which again reflects the strong Ni--O hybridization. These fluorescence lines also show very small polarization dependence, consistent with nearly equally distributed holes in $3d_{x^2-y^2}$ and $3d_{3z^2-r^2}$ orbitals. Such polarization-dependent fluorescence signals have previously been well captured in \ce{La4Ni3O10} using orbital-resolved density of states from \gls*{DFT} calculations \cite{Norman2023Orbital}.

\section{Magnetic excitations}

\begin{figure}
\includegraphics{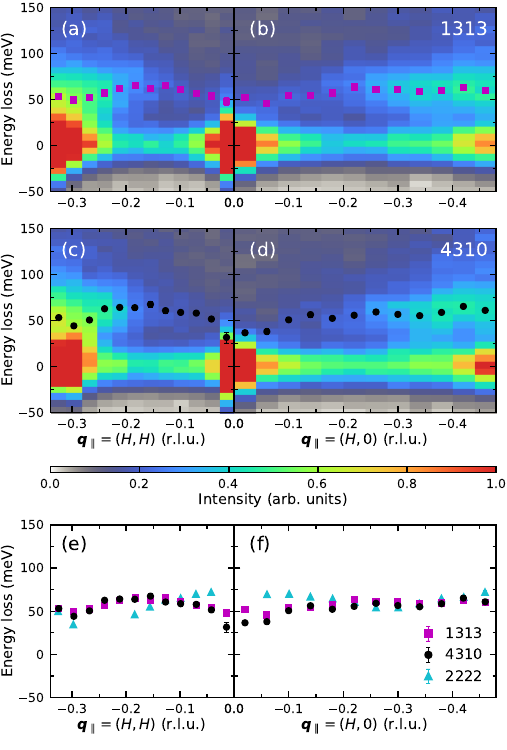}
\caption{In-plane magnetic-excitation dispersion of \ce{La4Ni3O10} and LNO-1313. (a)--(d) Ni $L_3$-edge \gls*{RIXS} intensity maps as a function of the in-plane momentum transfer $\bm{q}_{\parallel}$ along the (a),(c) $(H,H)$ and (b),(d) $(H,0)$ directions, respectively, with a focus on the low-energy magnetic excitations. The overlaid black dots (for \ce{La4Ni3O10}) and magenta squares (for LNO-1313) are extracted magnetic-excitation energies from the fits described in Supplemental Material Sec.~S6. (e),(f) Comparison of the fitted magnetic-excitation dispersion in these two materials. All the measurements were taken at $T=80$~K using $\pi$-polarized incident x-rays at an incident energy of $\sim 852.9$~eV to maximize the magnetic-excitation intensities. The scattering angle was fixed at $2\Theta = 150^\circ$, so the out-of-plane momentum component also varies for different $\bm{q}_{\parallel}$ positions. Error bars represent one standard deviation and include the combined uncertainties from the fitted magnetic-excitation peak position and energy zero. We also overlay the data for LNO-2222 extracted from Ref.~\cite{Chen2024Electronic} for direct comparison.
}
\label{fig:th-dep}
\end{figure}

\begin{figure}
\includegraphics{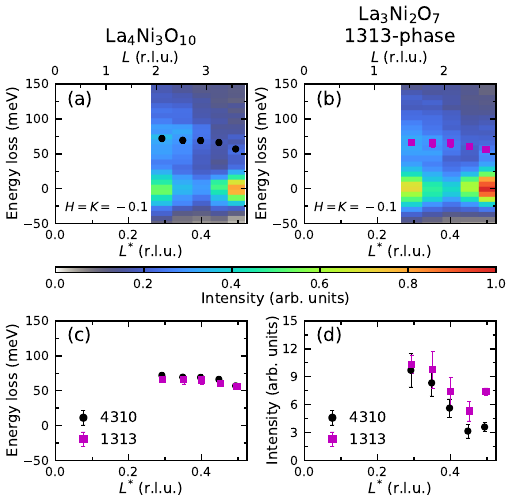}
\caption{Out-of-plane magnetic-excitation dispersion of \ce{La4Ni3O10} and LNO-1313. (a),(b) Ni $L_3$-edge \gls*{RIXS} intensity maps as a function of the out-of-plane momentum transfer with a focus on the low-energy magnetic excitations. The overlaid black dots (for \ce{La4Ni3O10}) and magenta squares (for LNO-1313) are extracted magnetic-excitation energies from the fits described in Supplemental Material Sec.~S6. Both the incident angle $\theta$ and scattering angle $2\Theta$ were adjusted to ensure a fixed $\bm{q}_{\parallel}=(-0.1, -0.1)$ r.l.u. The coordinate $L$ is based on the $c$-axis lattice constant, whereas $L^*=Ld/c$ is based on the interlayer separation of \ce{NiO2} planes $d\approx 3.9$~\AA{}. (c),(d) Comparison of the (c) fitted magnetic-excitation energies and (d) integrated intensities as a function of $L^*$ in these two materials. All the measurements were taken at $T=22$~K using $\pi$-polarized incident x-rays at an incident energy of $\sim 852.9$~eV to maximize the magnetic-excitation intensities. Error bars represent one standard deviation, and those for magnetic-excitation energies include combined uncertainties from the fitted peak position and energy zero.
}
\label{fig:L-dep}
\end{figure}

We next focus on the low-energy region of \gls*{RIXS}, using high-resolution data to investigate magnetic excitations in these materials. Ni $L_3$-edge \gls*{RIXS} has been extensively employed to study magnetic excitations in nickelates of various stoichiometries, such as 214 (denoting $R_2$NiO$_4$, and similarly for the following) \cite{Fabbris2017Doping}, 113 \cite{Lu2018SiteSelective}, 438 \cite{Lin2021Strong}, 112 \cite{Lu2021Magnetic, Hayashida2024Investigation, Gao2024Magnetic, Fan2024Capping, Worm2024Spin, Rossi2024Universal}, and, more recently, LNO-2222 \cite{Chen2024Electronic, Zhong2026Spin, Zhong2026Doping, Zhang2026Interlayer, Chen2026Orbital, Li2026Interlayer} and \ce{Nd4Ni3O10} \cite{TenHuisen2025Magnetic}. We use $\pi$-polarized incident x-rays with energy tuned to the first resonance peak at the Ni $L_3$-edge to optimize the intensity of our observed low-energy features. The in-plane momentum-dependent \gls*{RIXS} spectra are obtained by varying the incident x-ray angle while keeping the scattering angle fixed. As shown in Fig.~\ref{fig:th-dep}, \ce{La4Ni3O10} and LNO-1313 exhibit remarkably similar low-energy excitations along the two high-symmetry directions in reciprocal space. The equivalent dispersion measured along the $(H,-H)$ direction agrees closely with that along $(H,H)$, consistent with similarly populated structural twin domains (see Supplemental Fig.~S5). The spectra comprise an elastic peak and a broad dispersive feature centered around $60$~meV. This inelastic signal has comparable energy and intensity to the previously identified magnetic excitations in LNO-2222 (see Supplemental Material Sec.~S5), for which the magnetic origin was corroborated by polarimetric analysis \cite{Chen2024Electronic}. Therefore, we attribute this feature primarily to magnetic excitations.

To further quantify the magnetic excitations, we use \gls*{DHO} functions to fit the data. We find that two \gls*{DHO} peaks are sufficient to describe the broad inelastic feature up to $400$~meV (see fitting details in Supplemental Material Sec.~S6 \cite{supp}): a dominant peak around $60$~meV corresponding to magnetic excitations and a higher-energy shoulder near $160$~meV attributed to higher-order magnetic excitations. As shown in Figs.~\ref{fig:th-dep}(e) and (f), the in-plane magnetic-excitation dispersions closely track each other in \ce{La4Ni3O10} and LNO-1313, with broadly similar intensity trends (see Supplemental Fig.~S7). Along the $(H,0)$ direction, the fitted magnetic-excitation peak becomes stronger and disperses slightly upward from the Brillouin-zone center, reaching a maximum of $\sim 65$~meV near $(-0.5, 0)$. These behaviors are opposite to the magnetic-excitation trends reported in LNO-2222 \cite{Chen2024Electronic}. Along the $(H,H)$ direction, we see a softening of the fitted magnetic-excitation peak near $(1/3, 1/3)$, accompanied by an especially strong elastic line at this momentum. These observations are consistent with the previously identified spin order at $(0.31, 0.31)$ in \ce{La4Ni3O10} \cite{Zhang2020Intertwined, Samarakoon2023Bootstrapped}, and suggest the emergence of a similar spin order near $(1/3, 1/3)$ in LNO-1313. No peak is observed at the LNO-2222 spin-order wavevector $(0.25, 0.25)$ \cite{Chen2024Electronic, Ren2025Resolving, Gupta2025Anisotropic}. Taken together, these observations provide the first momentum-resolved \gls*{RIXS} evidence that the magnetic correlations in LNO-1313 more closely resemble those of trilayer \ce{La4Ni3O10} than those of bilayer LNO-2222. Compared with recent \gls*{RIXS} measurements in thin-film \ce{Nd4Ni3O10} \cite{TenHuisen2025Magnetic}, our magnetic-excitation energy is $\sim 10$~meV higher. The reduced magnetic-excitation bandwidth in thin films may result from the small tensile strain imposed by the substrate, which has been shown to effectively lower the corresponding energy in thin-film LNO-2222 \cite{Zhong2026Spin}.

As discussed in the previous section, there is substantial hybridization between inner apical O $2p_z$ and Ni $3d_{3z^2-r^2}$ orbitals. Consequently, it is reasonable to expect strong out-of-plane superexchange coupling. We therefore performed out-of-plane-dependent \gls*{RIXS} measurements at a fixed in-plane momentum of $(-0.1, -0.1)$ to probe possible dispersion along the out-of-plane direction. Indeed, as shown in Fig.~\ref{fig:L-dep}, the fitted magnetic-excitation peak gradually shifts to higher energy with enhanced intensity at smaller $L^*$ values. The magnetic-excitation energies and intensities remain very similar between \ce{La4Ni3O10} and LNO-1313, hinting at similar spin-exchange interactions in these two materials.

It is worth noting that even though only a single \gls*{DHO} function is used to fit the magnetic-excitation peak here, multiple non-degenerate magnetic-excitation modes are expected due to the large magnetic unit cell for \ce{La4Ni3O10} and LNO-1313. This is evident in the model calculations as detailed below. The observed peak thus represents an average over all branches, weighted by their respective cross-sections.

\begin{figure*}
\includegraphics[width=\textwidth]{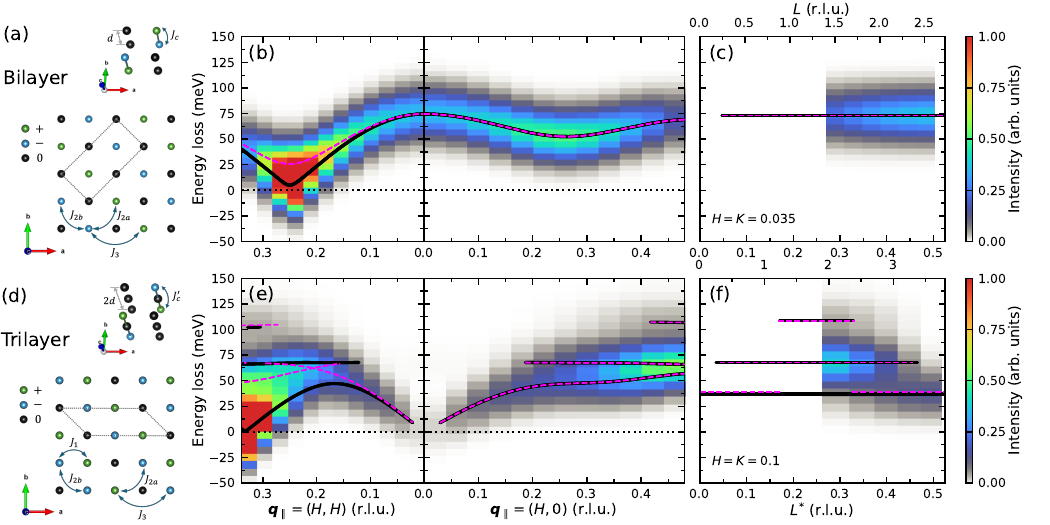}
\caption{Simulated magnetic-excitation spectra for a bilayer model of LNO-2222 and a trilayer model of \ce{La4Ni3O10} and LNO-1313. (a),(d) Schematic illustration of the magnetic structures and relevant exchange interactions for the bilayer and trilayer models, respectively. The green, blue, and black spheres represent spin-up, spin-down, and spinless Ni sites, respectively. For each panel, the bottom part displays one representative layer in the $ab$ plane and the top part displays the structure along the out-of-plane $c$ axis. Dotted lines illustrate the in-plane magnetic unit cells. (b),(e), Calculated magnetic-excitation spectra projected onto the in-plane high-symmetry $(H,H)$ and $(H,0)$ directions for the bilayer and trilayer models, respectively, with $L$ following the same trajectory as in Figs.~\ref{fig:th-dep}(a)--\ref{fig:th-dep}(d). (c),(f), Calculated magnetic-excitation spectra as a function of out-of-plane momentum at a fixed in-plane momentum of $(0.035, 0.035)$ for the bilayer and $(0.1, 0.1)$ for the trilayer. For both panels, the top axis shows the out-of-plane momentum in units related to the unit cell length with $c=20.45$~\AA{} in (c) and $c = 27.96$~\AA{} for (f). The bottom axis plots the same momentum in units related to the reciprocal \ce{NiO2}-layer separation of $d\approx 3.9$~\AA{}, i.e. $L^*=L d/c$. For LNO-2222, the simulation is the best fit to the data from Ref.~\cite{Chen2024Electronic}, while for \ce{La4Ni3O10}, the simulation is the best fit to the data in this study. The dispersion for LNO-1313 is the same within the uncertainties. Each intensity map is an incoherent average of contributions from two orthogonal magnetic stripe domains with equal populations; the maps are broadened by the instrument resolution and plotted on the same $\bm{Q}$ grid as the experimental data. The curves show the visible magnetic-excitation bands from each domain: the solid black lines correspond to the domain with the spin-order wavevector along $(H,H)$, while the dashed magenta lines correspond to the other domain, rotated by $90^{\circ}$. Horizontal dotted lines in (b) and (e) are included to indicate the zero-energy-loss position.
}
\label{fig:SW_calculations}
\end{figure*}

\section{Modeling of magnetic excitations}

To gain more insight into the ground-state magnetic structures and spin-exchange interactions, we model the measured in-plane and out-of-plane magnetic-excitation spectra using a Heisenberg Hamiltonian with the exchange paths and magnetic unit cells illustrated in Figs.~\ref{fig:SW_calculations}(a) and (d).  Given the similarity in the magnetic dispersions for \ce{La4Ni3O10} and LNO-1313, we construct one trilayer model for these materials, which we compare to a separate bilayer model for LNO-2222. Motivated by strong Hund’s coupling and the successful description of the dispersion and spectral-weight modulation as shown below, we use the Heisenberg framework as an effective theory of the magnetic sector. This approach provides a physically transparent means of quantifying the dominant exchange interactions and, crucially, the interlayer quantum-entangled states. The fitted couplings should be regarded as renormalized parameters that incorporate the influence of electronic correlations and itinerancy at an effective level. The detailed influence of band physics could be explored through future complementary \gls*{RPA}-type calculations based on a correlated band structure.

For the bilayer LNO-2222 case, we consider the \gls*{RIXS} data from Ref.~\cite{Chen2024Electronic}, including four in-plane high-symmetry directions as well as the $L^*$ dependence measured at a fixed in-plane position of $\bm{q}_{\parallel}=(0.035, 0.035)$ r.l.u. Because of the limited energy resolution in modern \gls*{RIXS}, only one branch was discernible in the data. However, a recent neutron scattering study \cite{Chen2026Nature} reported two non-degenerate modes with nonzero spin-gap values of $\sim5$~meV and $25$~meV at the spin order wavevector of $(0.25, 0.25)$. Although no clear shift of the magnetic-excitation peak position was detected in the $L^*$-dependent data, the spectral-weight modulation is considerable and consistent with the recent neutron results \cite{Chen2026Nature}, indicating antiferromagnetic interlayer coupling. In our model, we jointly optimize all in-plane and $L^*$-dependent \gls*{RIXS} datasets while imposing explicit constraints on the gap sizes based on the neutron results. We assume a spin-charge stripe pattern with high-spin $S=1$ moments on the magnetic Ni sites. The validity of this assumption is discussed further in Sec.~\ref{sec:discussion}. Since the formation of stripes naturally breaks the four-fold symmetry of the underlying square lattice, two magnetic twin domains with orthogonal orientations are possible. We consequently include magnetic domain-averaging effects in our modeling and find that our data are consistent with equal domain population~\footnote{We additionally tested and found that the conclusions we obtain are not highly sensitive to assumptions about the domain population.}. 

The key methodological advance in our modeling is the adoption of a generalized spin-wave calculation built on an entangled-unit formalism \cite{Dahlbom2024Classical, Dahlbom2025Sunny}, rather than the conventional linear spin-wave theory that has been employed in recent studies \cite{Chen2024Electronic, Zhong2026Spin, Zhong2026Doping, Zhang2026Interlayer, Chen2026Orbital, Chen2026Nature}.  This formalism starts by treating spins connected via $c$-axis bonds in terms of their full internal Hilbert subspace, allowing the presence of either quantum or classical states along these bonds. By contrast, conventional linear spin-wave theory expands about a magnetically ordered state and therefore cannot describe the singlet ground state or its associated gapped triplet excitations that have been studied in prior work on strongly coupled bilayer Heisenberg systems \cite{Ganesh2011Neel, Lohofer2015Dynamical}. Details of the generalized calculation are given in Supplemental Material Sec.~S7. Compared with the traditional spin-wave scenario reported in Ref.~\cite{Chen2024Electronic}, this entangled-unit formalism performs better, capturing not only the magnetic-excitation dispersion but also the momentum dependence of the spectral weight, as shown in Figs.~\ref{fig:SW_calculations}(b),\ref{fig:SW_calculations}(c) and Supplemental Fig.~S12. Most importantly, the extracted coupling parameters indicate a strong antiferromagnetic interlayer coupling, about an order of magnitude larger than the intralayer couplings, thereby supporting a dimer scenario for the ground state and low-energy excitations. The obtained optimal exchange-coupling parameters for LNO-2222 are $J_{2a}=1.23(4)$~meV, $J_{2b}=0.1(2)$~meV, $J_{3}=4.3(1)$~meV, and $J_c=52.6(6)$~meV. Unequal $J_{2a}$ and $J_{2b}$ are important for breaking the four-fold symmetry of the spin Hamiltonian and producing the two non-degenerate bands from the twin domains. Including both $J_{2}$ and $J_{3}$ is necessary to generate the anisotropic in-plane elliptical rings observed in neutron scattering \cite{Chen2026Nature}. In this parameter regime, the spectrum on these energy and resolution scales is dominated by one mode per magnetic domain. Within the constrained fit, the dimer scenario accommodates a triplon gap at the spin order wavevector without resorting to additional terms in the Hamiltonian (such as the single-ion anisotropy used in Ref.~\cite{Chen2026Nature}).

Having established the dimer scenario as an effective description of the magnetic excitations in LNO-2222, we extend this formalism to the trilayer case. Because LNO-1313 and \ce{La4Ni3O10} exhibit very similar spin ordering wavevectors, we use \ce{La4Ni3O10} as a representative example, for which more extensive literature is available. An intertwined spin and charge ordered state has been identified in \ce{La4Ni3O10} by x-ray and neutron diffraction measurements \cite{Zhang2020Intertwined, Samarakoon2023Bootstrapped}, which indicate a non-magnetic inner layer sandwiched between two outer layers with out-of-phase spin order. This M/0/M-type ground-state configuration, with high-spin $S=1$ moments residing on the outer layers, was also identified as the ground state of recent DFT calculations \cite{LaBollita2024ElectronicStructure}. We therefore adopt this spin configuration, as displayed in Fig.~\ref{fig:SW_calculations}(d), and include the two spins on the outer layers linked by the exchange coupling $J_c'$ within this effective model, which we solve while including the full Hilbert space of effective states between these pairs of spins. For the in-plane spin configuration, we assume an up-down-spinless pattern with a spin ordering wavevector of $(1/3, 1/3)$, close to the observed wavevector of $(0.31, 0.31)$ for \ce{La4Ni3O10} \cite{Zhang2020Intertwined}. This choice of magnetic ground state greatly reduces the computational complexity owing to its relatively small magnetic unit cell. We perform a joint optimization of the in-plane and out-of-plane momentum-dependent datasets and consider the presence of two twin stripe domains. Compared with the bilayer case for LNO-2222, the assumed spin ordering pattern here, with $J_1$ and $J_3$ alone, already breaks the four-fold symmetry. We therefore set both $J_{2a}$ and $J_{2b}$ to zero for simplicity. This $J_1$-$J_3$-$J_c'$ model reproduces the measured dispersion and the spectral-weight modulation, as demonstrated in Figs.~\ref{fig:SW_calculations}(e) and (f), with best-fit parameters $J_1=16(1)$~meV, $J_3=5.0(5)$~meV, and $J_c'=22(1)$~meV. Within the assumptions of the model, the $L^*$ dependence supports an M/0/M configuration, with antiparallel spins on the two outer layers and a non-magnetic node on the inner layer. Introducing a sizable magnetic moment on the inner layer would select an up-down-up type of ground-state spin configuration and double the spectral-weight modulation periodicity along $L^*$, which would be inconsistent with the data. The effective interlayer exchange interaction $J_c'$ determined from this model remains substantial compared with the in-plane interactions, despite the longer interlayer distance across the non-magnetic inner layer.

\section{Lattice excitations}

\begin{figure}
\includegraphics{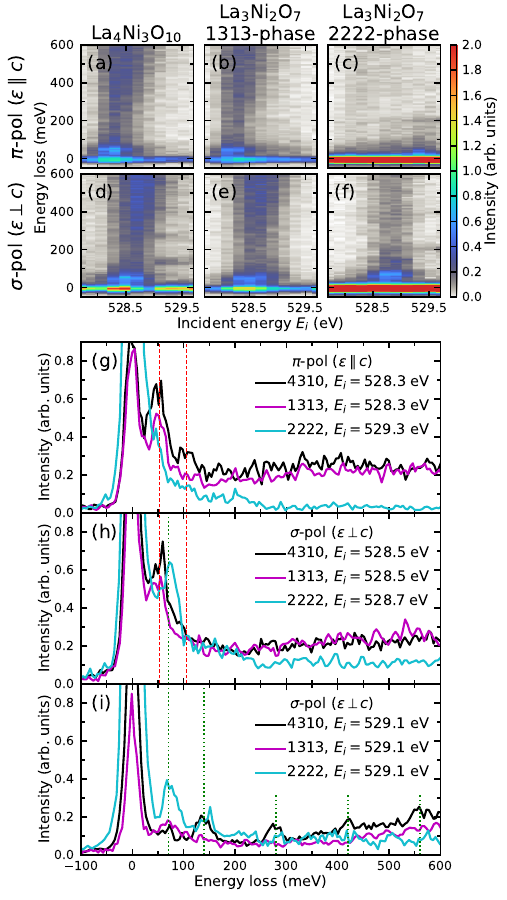}
\caption{Comparison of phonons in \ce{La4Ni3O10} and the two polymorphs of \ce{La3Ni2O7}. (a)--(f) O $K$-edge \gls*{RIXS} intensity maps as a function of incident x-ray energy. These maps are zoomed-in views of the data in Figs.~\ref{fig:energy_map}(i)--\ref{fig:energy_map}(n), emphasizing the low-energy phonon features. Panels (g)--(i) display several representative \gls*{RIXS} spectra with red and green vertical lines marking multiples of 53 and 70~meV, respectively, based on \ce{La4Ni3O10} and LNO-1313. The data were taken at 22~K with an incident angle of $\theta = 15^\circ$ and scattering angle of $2\Theta = 150^\circ$ in the ($HHL$) scattering plane. The polarization of $\sigma$($\pi$)-polarized incident x-rays is perpendicular (approximately parallel) to the sample $c$-axis.
}
\label{fig:phonons}
\end{figure}

In addition to probing electronic and magnetic excitations, \gls*{RIXS} is also sensitive to lattice excitations, i.e., phonons, and serves as an increasingly powerful tool for probing the strength of \gls*{EPC} \cite{Yavas2010Observation, Ament2011Determining, Meyers2018Decoupling, Dashwood2021Probing, Devereaux2016Directly, Braicovich2020Determining, Li2020Multiorbital, Peng2020Enhanced, Peng2022Doping, Thomas2025Theory, Hong2025Dominant}. In particular, out-of-plane bond-buckling and in-plane bond-stretching phonons, involving oxygen displacements, are the two most commonly observed lattice-vibration modes in prior \gls*{RIXS} studies of cuprates \cite{Braicovich2020Determining, Li2020Multiorbital, Peng2020Enhanced, Peng2022Doping}. O $K$-edge \gls*{RIXS} is uniquely suited for such measurements owing to its direct sensitivity to oxygen sites, longer core-hole lifetime, and the absence of overlapping magnetic-excitation peaks. As shown in Fig.~\ref{fig:phonons}, our O $K$-edge measurements identify these two phonon modes in nickelates for the first time. The incident-energy-dependent \gls*{RIXS} maps in panels (a)--(f) are magnified views of Figs.~\ref{fig:energy_map}(i)--\ref{fig:energy_map}(n), highlighting the low-energy excitations. The broad peak near $400$~meV in \ce{La4Ni3O10} and LNO-1313 is consistent with the previously discussed feature in the Ni $L_3$-edge data, suggesting increased metallicity. Clear, strong peaks are present in all three materials below $\sim 75$~meV, consistent with the phonon energy scales in these materials predicted by recent \gls*{DFT} calculations \cite{Zhang2024Structural, LaBollita2024ElectronicStructure}. We therefore attribute these sharp features to phonons.

To facilitate comparison of the phonon peaks across the three materials, we plot several representative \gls*{RIXS} spectra in Figs.~\ref{fig:phonons}(g)--\ref{fig:phonons}(i). In the $\pi$-polarized channel at the main resonance, all three compounds exhibit a strong phonon peak at $53$~meV and a weaker second harmonic [indicated by the two red dashed lines in panel (g)]. We assign these peaks to the fundamental and second harmonic of the bond-buckling mode involving out-of-plane oxygen vibrations. The fundamental energy is comparable to that of the corresponding mode in cuprates \cite{Devereaux2016Directly, Braicovich2020Determining, Li2020Multiorbital, Peng2022Doping}. This mode is stronger in \ce{La4Ni3O10} and LNO-1313 than in LNO-2222, indicating stronger \gls*{EPC} to this phonon branch. Another higher-energy peak emerges near $70$~meV in the $\sigma$-polarized spectra at the main resonant energy [panel (h)], as marked by the green dotted line. The energy of this peak is consistent with that of the in-plane bond-stretching phonon in cuprates \cite{Devereaux2016Directly, Braicovich2020Determining, Li2020Multiorbital, Peng2020Enhanced, Peng2022Doping, Hong2025Dominant}. This bond-stretching phonon peak is significantly stronger in LNO-2222 and exhibits a subtle blue shift of $\sim 3$~meV relative to the peaks in \ce{La4Ni3O10} and LNO-1313, as seen more clearly in the post-edge spectra shown in panel (i). Surprisingly, the post-edge spectrum of \ce{La4Ni3O10} also contains a ladder of peaks located at energies consistent with even multiples of the phonon energy. 
These peaks extend up to eight times the energy of the bond-stretching phonon. Such high-order phonon overtones are rare and are associated with very strong \gls*{EPC}, with a recent example being an apical-oxygen phonon mode in a trilayer cuprate \cite{Hong2025Dominant}. Within \gls*{EPC} frameworks, an unusual aspect of the data is the absence of peaks corresponding to odd orders of the phonon energy (except for the first-order one, which is also relatively weak). Such phonon structures are not captured by current theoretical models for phonons in \gls*{RIXS}~\cite{Ament2011Determining, Geondzhian2020generalization, Thomas2025Theory} and require additional modeling to be fully understood. 

\section{Discussion and conclusions}
\label{sec:discussion}

The central experimental contribution of this work is the first comprehensive \gls*{RIXS} characterization of LNO-1313, spanning its electronic, magnetic, and lattice excitations and benchmarking them directly against bilayer LNO-2222 and trilayer \ce{La4Ni3O10}. Our results demonstrate a strong similarity between bulk \ce{La4Ni3O10} and LNO-1313 at ambient pressure in terms of their electronic, magnetic, and lattice excitations, despite their different layer stackings and nominal Ni valences. A simple superposition of monolayer \ce{La2NiO4} and trilayer \ce{La4Ni3O10} motifs cannot account for the observed spectra of LNO-1313 (see Supplemental Fig.~S4), suggesting that the monolayer block in LNO-1313 is spectroscopically silent. Recent \gls*{DFT} plus \gls*{DMFT} studies suggest that the monolayer in LNO-1313 at ambient pressure is in a $d^8$ Mott-insulating state with a large energy gap of $\sim 4$~eV \cite{Lechermann2024Electronic, LaBollita2024Electronic}. This layer distinction could potentially explain the dominant role of the trilayer motif in shaping the low-energy physics in LNO-1313. Hence, the observed similarity reflects the strong Hund's coupling for the Ni sites, which favors the $d^8$ high-spin configuration and pushes any additional holes to the ligand sites.

In modeling the magnetic interactions in these materials, we have assumed a spin-charge stripe for the ground state \cite{Chen2024Electronic, Yashima2025Microscopic, Chen2026Nature}. Such an assumption is one of two leading experimental proposals, the other one being a double spin stripe model \cite{Chen2024Electronic, Ren2025Resolving, Gupta2025Anisotropic}. Both models are compatible with the observed in-plane spin-order wavevector of $(0.25, 0.25)$ reported by neutron scattering \cite{Plokhikh2026Unraveling, Chen2026Nature} and resonant x-ray diffraction \cite{Ren2025Resolving, Gupta2025Anisotropic} experiments. Recent nuclear quadrupole resonance (NQR) \cite{Yashima2025Microscopic} and muon-spin rotation/relaxation ($\mu$SR) \cite{Chen2024Evidence, Khasanov2025Pressureenhanced} measurements have provided direct evidence for two magnetically inequivalent Ni sites. Together with recent inelastic magnetic neutron scattering results \cite{Chen2026Nature}, these findings favor the spin-charge stripe model. In terms of the out-of-plane spin-order wavevector $q_L$, however, a discrepancy remains between $q_L = 1/2$ and $q_L = 0$ \cite{Gupta2025Anisotropic, Plokhikh2026Unraveling, Chen2026Nature}, which may arise from stacking faults between neighboring bilayer blocks. The short correlation length along $L$, observed in neutron scattering \cite{Chen2026Nature} and resonant x-ray diffraction \cite{Gupta2025Anisotropic} experiments, together with the negligible inter-bilayer coupling, suggests that such stacking faults can form easily. Consequently, the out-of-plane energy dispersion of specific magnetic modes is expected to be negligible. In contrast, the interlayer coupling within each bilayer block is strong and produces a pronounced spectral-weight modulation, as demonstrated by the inelastic neutron scattering \cite{Chen2026Nature} and \gls*{RIXS} \cite{Chen2024Electronic} measurements. This strong antiferromagnetic interlayer coupling, combined with the in-plane spin-charge stripe modulation, provides the basis for our dimer scenario for the low-energy spin physics of this material. In the strong-coupling limit, two neighboring magnetic Ni sites within a bilayer form a spin singlet, and triplon excitations of these dimers propagate through the lattice via in-plane exchange couplings. More precisely, the magnetic excitations in LNO-2222 are better described as triplons rather than as conventional magnons. This dimer picture also naturally accounts for the small ordered moment and large fluctuating moment observed in recent inelastic neutron scattering measurements \cite{Chen2026Nature}, because the singlet has nominally $S=0$ total spin. The observed static ordered moment may arise from triplon condensation or from an admixture of triplet character induced by interdimer coupling. Several works have suggested that the singlet-triplet excitations may have a key role in superconductivity in this material family \cite{Khaliullin2026orbital, Maier2026interlayer, Labollita2026squeezing}.

A possible microscopic picture of this stripe pattern is a spatial alternation between two types of Ni dimers: $d^8/d^8$ dimers forming the magnetic stripes and $d^8\underline{L}/d^8\underline{L}$ dimers forming the nominally spinless stripes. Because of strong Hund’s coupling, Ni sites tend to favor a $d^8$ electronic configuration with comparable occupation of the $d_{x^2-y^2}$ and $d_{3z^2-r^2}$ orbitals, resulting in an $S=1$ high-spin state. The two Ni spins are bridged by the nearly fully occupied inner apical oxygen $p_z$ orbital, giving rise to strong antiferromagnetic superexchange within the bilayer. In the neighboring dimer, the $d^8\underline{L}/d^8\underline{L}$ electronic configuration contains self-doped ligand holes predominantly on the planar oxygen $p_{x,y}$ orbitals. These holes tend to form Zhang-Rice-like singlet states with the in-plane Ni orbitals, leaving an effective low-spin state on each Ni site. The two low-spin Ni sites may further couple into a total-spin $S=0$ state, providing a possible microscopic origin for the spinless stripes.

Compared with the strong-dimer-limit picture proposed for bilayer LNO-2222, the trilayer nickelates, including LNO-1313 and \ce{La4Ni3O10}, may be better viewed as lying in an intermediate-coupling regime. The interlayer exchange remains strong, but the in-plane interactions are also enhanced. In particular, with $J_1/J_c^\prime\approx0.7$, the trilayer lies outside of the strong-rung limit, where neither an expansion about ordered moments (a traditional spin-wave calculation) nor a strong-dimer expansion is fully controlled \cite{Lohofer2015Dynamical, Ganesh2011Neel, zhang2025large}. The resulting fitted parameters should accordingly be regarded as effective values with correspondingly larger uncertainty. Nonetheless, a number of observations continue to support the dimer picture. Infrared spectroscopy shows that density-wave order in \ce{La4Ni3O10} strongly suppresses the out-of-plane conductivity through a redistribution of Ni $3d_{3z^2-r^2}$ occupation, producing an effective electronic decoupling of the layers \cite{Guan2026Electronic}. This picture is consistent with the observed weak orbital polarization, which suggests that holes are more uniformly distributed between apical and planar oxygen sites. Our $L^*$-dependent data provide evidence, from the inelastic scattering channel, for an M/0/M spin configuration with a non-magnetic inner layer. This is also consistent with a recent \gls*{ARPES} report that found stronger coupling between the two outer layers than between the outer and inner layers in explaining the observed electronic band structure in LNO-1313 \cite{Au-Yeung2026Oxygencentred}. While the fitting quality alone cannot definitively establish that a dimer scenario is superior to a conventional spin-wave scenario, the considerations above suggest that a dimer-based description is more theoretically appropriate and physically motivated, and may better capture the enhanced quantum fluctuations in this intermediate-coupling regime.

Phenomenologically, LNO-1313 and \ce{La4Ni3O10} show weakened orbital polarization and slightly reduced magnetic-excitation energies compared with bulk LNO-2222 \cite{Chen2024Electronic}. Our magnetic-excitation simulations also reveal that the effective out-of-plane exchange-coupling parameter is reduced by more than a factor of two, accompanied by enhanced in-plane exchange interactions. Therefore, we speculate that the higher tendency toward superconductivity of the LNO-2222 structure may be related to its enhanced interlayer exchange interaction pertinent to its bilayer structure. 

Our O $K$-edge data show rather similar one-phonon energies and intensities in \ce{La4Ni3O10} and LNO-1313. We observe bond-buckling and bond-stretching modes in all three compounds. The appearance of second-order peaks for both modes, particularly the unusually pronounced bond-stretching phonon overtones in \ce{La4Ni3O10}, indicates strong \gls*{EPC}, which might be relevant to superconductivity in these materials. The energy of the bond-stretching mode ($\sim 70$~meV) is also close to the kink feature observed for the $\alpha$ band (with leading $3d_{x^2-y^2}$ character) in \gls*{ARPES} \cite{Wang2026Electronic}, pointing to an \gls*{EPC} origin. However, the stronger \gls*{EPC} in trilayer \ce{La4Ni3O10}, despite its lower superconducting $T_\mathrm{c}$, argues against a simple picture in which \gls*{EPC} is the dominant mechanism for the superconductivity. 
Determining whether phonons play a secondary role will require more detailed mapping of the \gls*{EPC} in momentum space~\cite{Johnston2010systematic}. A recent momentum-resolved inelastic x-ray scattering study likewise found no appreciable phonon softening near the charge ordering wavevector in trilayer nickelates, indicating that their intertwined density-wave order is primarily spin driven rather than lattice driven \cite{Jia2026Lattice}.

In summary, our \gls*{RIXS} measurements provide a comparative view of the electronic, magnetic, and lattice excitations in bilayer LNO-2222, alternating monolayer–trilayer LNO-1313, and trilayer \ce{La4Ni3O10}. The bilayer compound exhibits excitation spectra that differ from those of the other two materials, whereas the layer-alternating and trilayer compounds show unexpectedly similar responses across all three excitation channels. These results indicate that layer architecture plays a central role in shaping the coupled electronic, magnetic, and lattice degrees of freedom of \gls*{RP} nickelates, providing constraints on their low-energy physics and the microscopic ingredients relevant to superconductivity.

\textit{Note added}: While preparing this manuscript, we became aware of studies of \ce{La4Ni3O10} reported in Refs.~\cite{Chen2026Dissecting, Chan2026Collective}, which report magnetic dispersions that are consistent with the data here.

The supporting data for the plots in this article are openly available from the Zenodo database \cite{repo}.

\begin{acknowledgments}
We thank Mike Norman for discussions. Work at Brookhaven and the University of Tennessee (RIXS measurements and interpretation) was supported by the U.S. Department of Energy, Office of Science, Office of Basic Energy Sciences, under Award No. DE-SC0022311. Work at Harvard was supported by the U.S. Department of Energy, Office of Science, Office of Basic Energy Sciences, under Award No. DE-SC0012704. Work at Argonne National Laboratory (crystal growth) was supported by the U.S.\ Department of Energy Office of Science, Basic Energy Sciences, Materials Science and Engineering Division. W.H.\ acknowledges additional support by the U.S. Department of Energy, Office of Science, Basic Energy Sciences, Materials Sciences and Engineering Division, under contract DE-AC02-76SF00515. This research used resources at the SIX beamline of the National Synchrotron Light Source II, a U.S. DOE Office of Science User Facility operated for the DOE Office of Science by Brookhaven National Laboratory under Contract No.~DE-SC0012704. The Flatiron Institute is a division of the Simons Foundation.
\end{acknowledgments}


\bibliography{refs}
\end{document}


\title{Supplemental Material: Layer Architecture Shapes Electronic, Magnetic, and Lattice Interactions in Ruddlesden-Popper Nickelates}

\renewcommand{\thesection}{S\arabic{section}}
\renewcommand{\thetable}{S\arabic{table}}
\renewcommand{\thefigure}{S\arabic{figure}}
\renewcommand{\theequation}{S\arabic{equation}}

\date{\today}

\maketitle

This document provides \gls*{XRD} data, polarization-dependent \gls*{XAS} data, an examination of the superposition assumption for LNO-1313, additional in-plane momentum-dependent \gls*{RIXS} spectra in the $(H,-H,L)$ plane for LNO-1313, a comparison of magnetic-excitation intensities, a detailed description of the fitting procedures for the \gls*{RIXS} data with line cuts showing the fits, and modeling of the magnetic excitations with additional simulations for LNO-2222.

\tableofcontents

\section{Laboratory XRD data}
To confirm the phase purity of the single crystals used in this study, we measured our samples using laboratory \gls*{XRD}.

\begin{figure}
\includegraphics{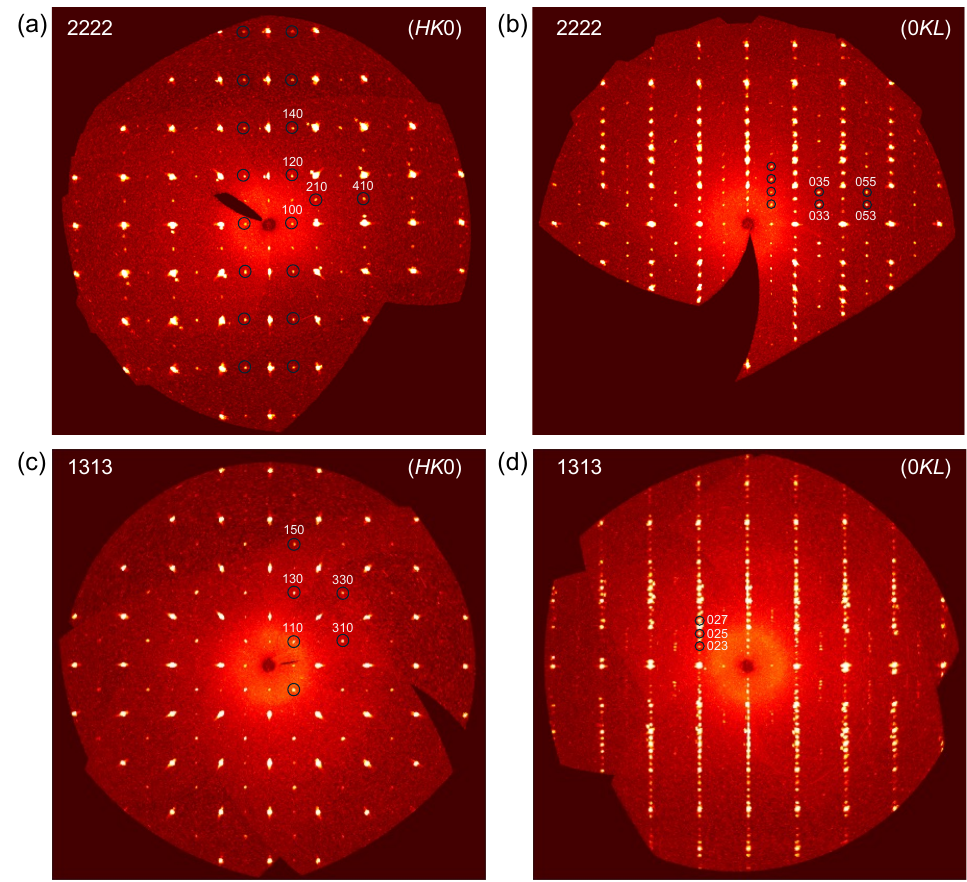}
\caption{Precession images of LNO-2222 and LNO-1313 crystals. (a),(b) $(HK0)$ and $(0KL)$, respectively, for LNO-2222. (c),(d) $(HK0)$ and $(0KL)$, respectively, for LNO-1313.}
\label{fig:SI_XRD_La327}
\end{figure}

The LNO-2222 and LNO-1313 crystals used in this study were approximately 100~$\mu$m in size. We therefore performed single-crystal \gls*{XRD} measurements and plot the resulting precession images in both the $(HK0)$ and $(0KL)$ planes in Fig.~\ref{fig:SI_XRD_La327}. The characteristic weak reflections marked by open circles provide clear fingerprints of the expected phase in each compound \cite{Chen2024Polymorphism}. For LNO-2222, the selection rules $K=2n$ in the $(HK0)$ plane and $K+L=2n$ in the $(0KL)$ plane are consistent with its $Amam$ space group. In contrast, the reflection patterns in LNO-1313 follow the $Cmmm$ space-group selection rules, i.e., $H+K=2n$ in the $(HK0)$ plane and $K+L=2n$ in the $(0KL)$ plane. For LNO-2222, additional reflections satisfying $H=2n$ are observed in the $(HK0)$ plane, indicating the presence of structural twinning. The violations of these reflection conditions near the origin likely arise from $\lambda/2$ contamination.

\begin{figure}
\includegraphics{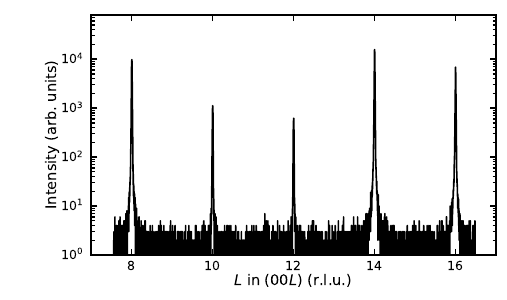}
\caption{$(00L)$ reflections of \ce{La4Ni3O10} measured by a $\theta$--2$\Theta$ \gls*{XRD} scan.}
\label{fig:SI_XRD_La4310}
\end{figure}

The \ce{La4Ni3O10} sample used in this study is approximately 2~mm in size and is too thick for single-crystal \gls*{XRD} in transmission, so we instead measured the $L$ dependence in reflection geometry. As shown in Fig.~\ref{fig:SI_XRD_La4310}, the positions and intensities of the ($00L$) peaks are consistent with the previously reported \ce{La4Ni3O10} structure \cite{Zhang2020High}, with no detectable impurity phases.

\section{Polarization-dependent XAS data}

\begin{figure}
\includegraphics{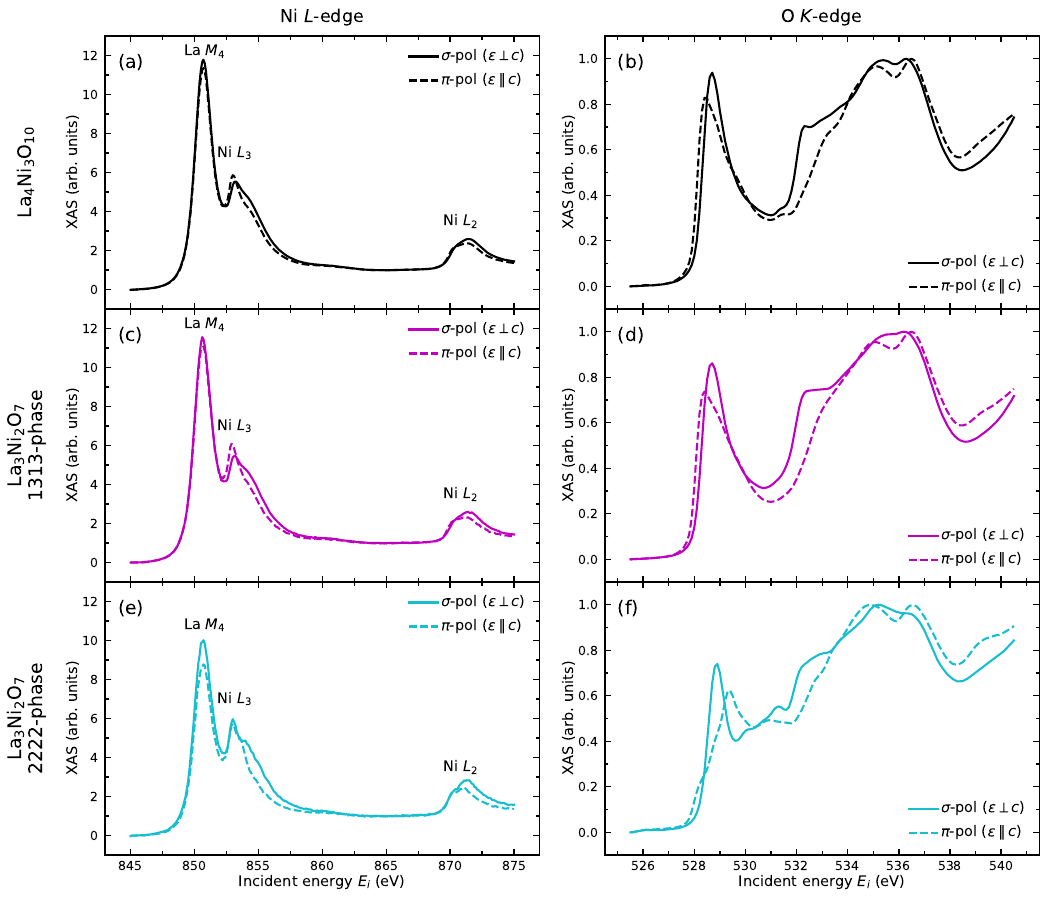}
\caption{Polarization-dependent \gls*{XAS} spectra of \ce{La4Ni3O10} and the two polymorphs of \ce{La3Ni2O7}, measured in total fluorescence yield mode. The data are the same as the \gls*{XAS} data shown in Fig.~1(a) and (b) and are replotted to highlight the linear dichroism of each material. The \gls*{XAS} $L$-edge data are normalized to the post-$L_3$-edge value in each spectrum. Because the O $K$-edge spectra span a much wider energy range, each spectrum is normalized to its maximum value. The Ni $L$-edge data were taken at 80~K with an incident angle of $\theta = 10^\circ$ and scattering angle of $2\Theta = 150^\circ$ in the ($H0L$) scattering plane, while the O $K$-edge data were taken at 22~K with an incident angle of $\theta = 15^\circ$ and scattering angle of $2\Theta = 150^\circ$ in the ($HHL$) scattering plane.}
\label{fig:SI_XAS}
\end{figure}

To provide a direct visualization of the linear dichroism in the studied materials, we present polarization-dependent \gls*{XAS} data in Fig.~\ref{fig:SI_XAS}. It is clear that LNO-2222 exhibits slightly stronger linear dichroism than the other two materials, particularly at the oxygen $K$-edge. Nonetheless, the linear dichroism in all three materials is significantly weaker than that observed in square-planar nickelates \cite{Zhang2017Large, Rossi2021Orbital} or cuprates \cite{Chen1992Outofplane}.

\section{Examination of the superposition assumption for LNO-1313}

\begin{figure}
\includegraphics{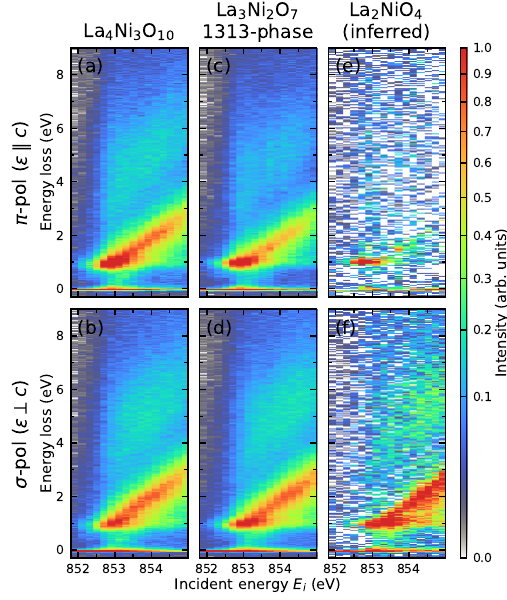}
\caption{Inferred Ni $L_3$-edge \gls*{RIXS} energy maps for single-layer \ce{La2NiO4}. (a)--(d) \gls*{RIXS} energy maps of \ce{La4Ni3O10} and LNO-1313, replotted from Fig.~1. (e),(f) Inferred \gls*{RIXS} energy maps of \ce{La2NiO4}, assuming LNO-1313 is a linear superposition of \ce{La4Ni3O10} and \ce{La2NiO4}. The evident mismatch between the inferred and previously measured \ce{La2NiO4} \gls*{RIXS} maps \cite{Fabbris2017Doping} thus invalidates this assumption.
}
\label{fig:SI_Inferred_214_maps}
\end{figure}

We examine the validity of the superposition assumption for LNO-1313, i.e., that the electronic structure of LNO-1313 can be regarded as a linear superposition of trilayer \ce{La4Ni3O10} and monolayer \ce{La2NiO4} blocks in a 3:1 ratio. As shown in Fig.~\ref{fig:SI_Inferred_214_maps}, the inferred Ni $L_3$-edge \gls*{RIXS} energy maps for \ce{La2NiO4} exhibit broad $dd$ excitation features, particularly in the $\sigma$-polarization channel, similar to those observed in \ce{La4Ni3O10} or LNO-1313. Such broad features differ markedly from the sharp, well-isolated peaks reported in previous \ce{La2NiO4} \gls*{RIXS} measurements \cite{Fabbris2017Doping}. The clear discrepancy therefore rules out the validity of the superposition assumption.

\section{Additional in-plane momentum-dependent RIXS spectra in the \texorpdfstring{$(H,-H,L)$}{(H,-H,L)} plane for LNO-1313}

\begin{figure}
\includegraphics{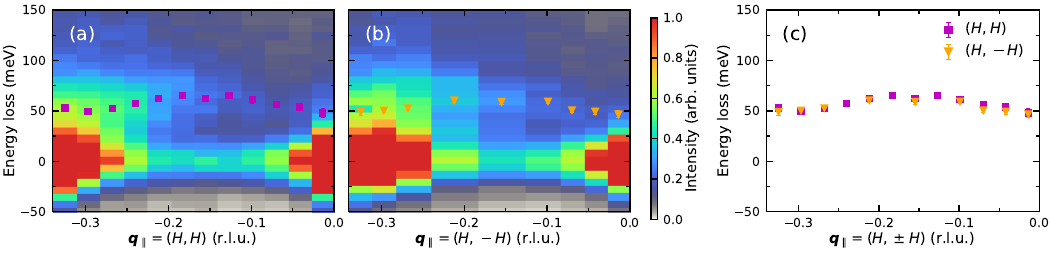}
\caption{Comparison of the in-plane magnetic-excitation dispersion in LNO-1313 along the $(H,H)$ and $(H,-H)$ directions. (a) Replotted Ni $L_3$-edge \gls*{RIXS} intensity map as a function of the in-plane momentum transfer $\bm{q}_{\parallel}$ along the $(H,H)$ direction from Fig.~2(a). (b) Ni $L_3$-edge \gls*{RIXS} intensity map as a function of the in-plane momentum transfer $\bm{q}_{\parallel}$ along the $(H,-H)$ direction. (c) Comparison of the fitted magnetic-excitation dispersion along the two directions. All measurements were taken at $T=80$~K using $\pi$-polarized incident x-rays at an incident energy of $\sim 852.9$~eV to maximize the magnetic-excitation intensities. The scattering angle was fixed at $2\Theta = 150^\circ$. Error bars represent one standard deviation and include the uncertainties in both the fitted peak position and energy zero.}
\label{fig:SI_HmH_magnon_comparison}
\end{figure}

We present the additional in-plane magnetic-excitation dispersion of LNO-1313 in the $(H,-H,L)$ scattering plane in Fig.~\ref{fig:SI_HmH_magnon_comparison}. The extracted dispersion agrees well with the results obtained in the $(H,H,L)$ plane. This is consistent with our hypothesis of equally populated structural twin domains.

\section{Comparison of magnetic-excitation intensities}
\begin{figure}
\includegraphics{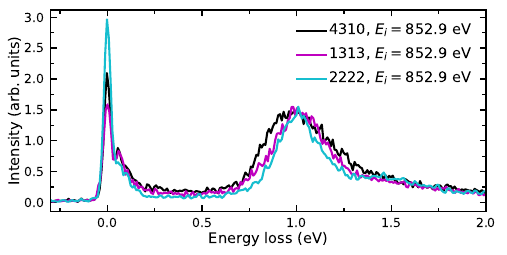}
\caption{Comparison of magnetic-excitation and $dd$-excitation intensities in \ce{La4Ni3O10} and the two polymorphs of \ce{La3Ni2O7}. All \gls*{RIXS} measurements were taken at $T=80$~K using $\pi$-polarized incident x-rays with an incident angle of $\theta = 10^\circ$ and scattering angle of $2\Theta = 150^\circ$ in the ($H0L$) scattering plane. The magnetic excitations and the $dd$ excitations near 1~eV show similar intensities in all three materials.
}
\label{fig:SI_RIXS_spectra_comparison}
\end{figure}

In Fig.~\ref{fig:SI_RIXS_spectra_comparison}, we show Ni $L_3$-edge \gls*{RIXS} spectra of the three studied compounds measured under the same experimental conditions. All three materials display comparable intensities for both the low-energy peak near 60--70~meV and the $dd$ excitations near 1~eV. Following the previous \gls*{RIXS} study that identified the low-energy feature in LNO-2222 as a magnetic excitation \cite{Chen2024Electronic}, we likewise assign this peak to magnetic excitations in \ce{La4Ni3O10} and LNO-1313.

\begin{figure}
\includegraphics{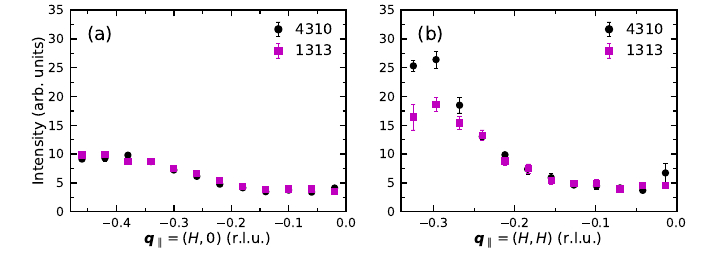}
\caption{Comparison of magnetic-excitation intensities between \ce{La4Ni3O10} and LNO-1313. The fitted oscillator strength of the magnetic-excitation peak obtained with the \gls*{DHO} model (see Section~\ref{sec:RIXS_fitting}) is plotted as a function of in-plane momentum transfer $\bm{q}_{\parallel}$ along the (a) $(H,0)$ and (b) $(H,H)$ directions.}
\label{fig:SI_thdep_intensity}
\end{figure}

Figure~\ref{fig:SI_thdep_intensity} compares the extracted magnetic-excitation intensities of \ce{La4Ni3O10} and LNO-1313 as a function of in-plane momentum transfer along two high-symmetry directions. The magnetic-excitation intensities closely match between \ce{La4Ni3O10} and LNO-1313, further suggesting similar magnetic excitations in the two materials.

\section{Fitting of the RIXS spectra}
\label{sec:RIXS_fitting}

\begin{figure}
\includegraphics{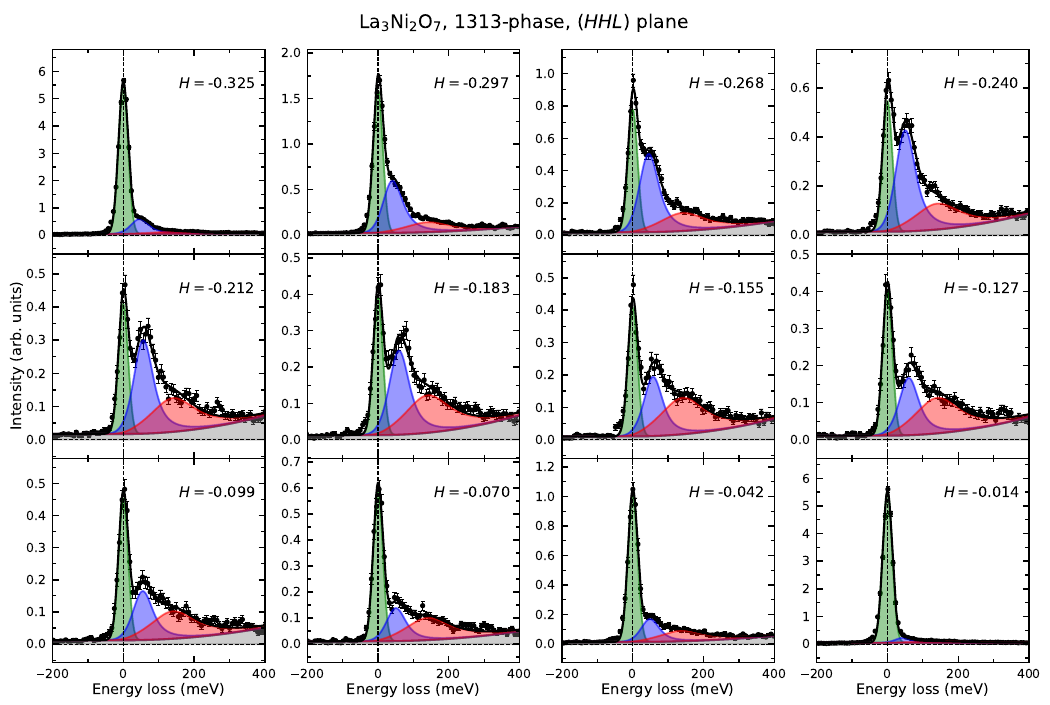}
\caption{\gls*{RIXS} spectra of LNO-1313 in the low-energy region, measured at different in-plane momentum transfers $H$ in the ($HHL$) scattering plane. The measurements were taken at $T=80$~K using $\pi$-polarized incident x-rays at an incident energy of $\sim 852.9$~eV. These data are the same as the intensity map shown in Fig.~2(a) and are provided to show the line cuts directly. Solid black lines are fits to the data, with shaded areas indicating contributions from different components (green for the elastic peak, blue for the magnetic-excitation peak, red for the higher-order magnetic-excitation peak, and gray for a quadratic background). Error bars represent one standard deviation.
}
\label{fig:SI_fits_1313_HHL}
\end{figure}

\begin{figure}
\includegraphics{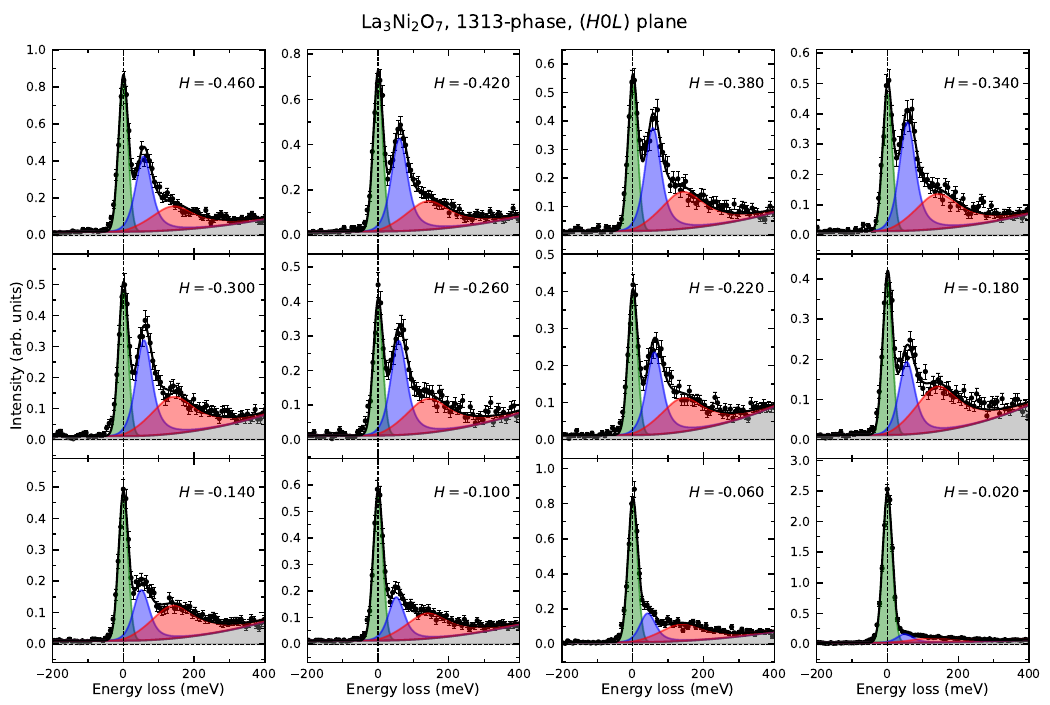}
\caption{\gls*{RIXS} spectra of LNO-1313 in the low-energy region, measured at different in-plane momentum transfers $H$ in the $(H0L)$ scattering plane. The measurements were taken at $T=80$~K using $\pi$-polarized incident x-rays at an incident energy of $\sim 852.9$~eV. These data are the same as the intensity map shown in Fig.~2(b) and are provided to show the line cuts directly. Solid black lines are fits to the data, with shaded areas indicating contributions from different components (green for the elastic peak, blue for the magnetic-excitation peak, red for the higher-order magnetic-excitation peak, and gray for a quadratic background). Error bars represent one standard deviation.
}
\label{fig:SI_fits_1313_H0L}
\end{figure}

\begin{figure}
\includegraphics{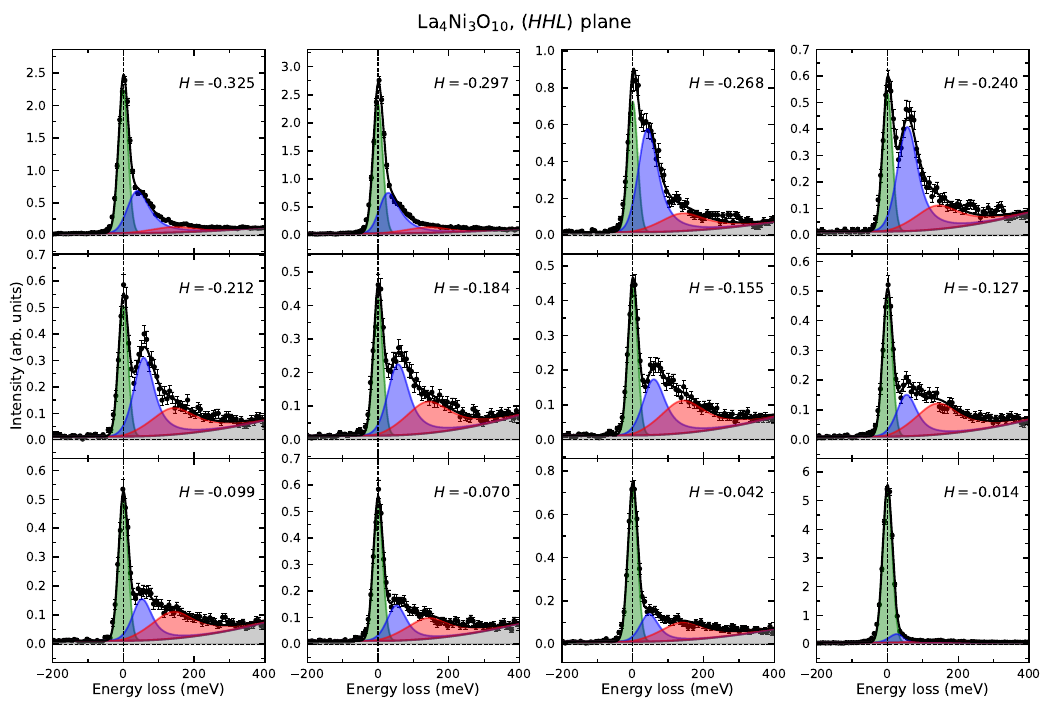}
\caption{\gls*{RIXS} spectra of \ce{La4Ni3O10} in the low-energy region, measured at different in-plane momentum transfers $H$ in the ($HHL$) scattering plane. The measurements were taken at $T=80$~K using $\pi$-polarized incident x-rays at an incident energy of $\sim 852.9$~eV. These data are the same as the intensity map shown in Fig.~2(c) and are provided to show the line cuts directly. Solid black lines are fits to the data, with shaded areas indicating contributions from different components (green for the elastic peak, blue for the magnetic-excitation peak, red for the higher-order magnetic-excitation feature, and gray for a quadratic background). Error bars represent one standard deviation.
}
\label{fig:SI_fits_4310_HHL}
\end{figure}

\begin{figure}
\includegraphics{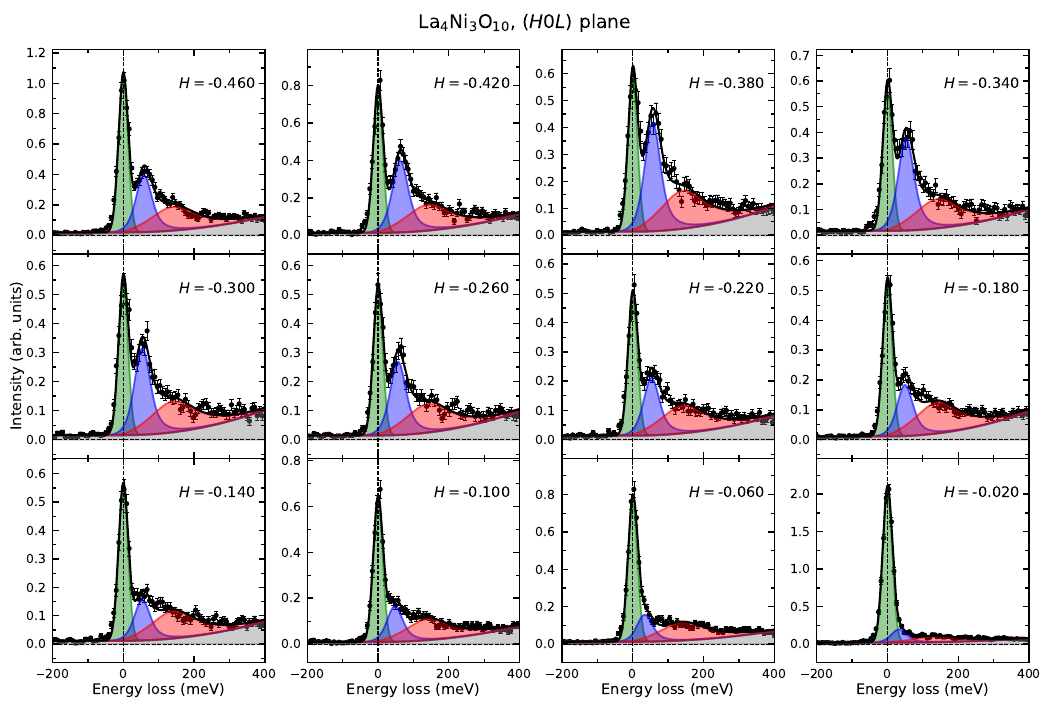}
\caption{\gls*{RIXS} spectra of \ce{La4Ni3O10} in the low-energy region, measured at different in-plane momentum transfers $H$ in the ($H0L$) scattering plane. The measurements were taken at $T=80$~K using $\pi$-polarized incident x-rays at an incident energy of $\sim 852.9$~eV. These data are the same as the intensity map shown in Fig.~2(d) and are provided to show the line cuts directly. Solid black lines are fits to the data, with shaded areas indicating contributions from different components (green for the elastic peak, blue for the magnetic-excitation peak, red for the higher-order magnetic-excitation feature, and gray for a quadratic background). Error bars represent one standard deviation.
}
\label{fig:SI_fits_4310_H0L}
\end{figure}

To quantify the magnetic-excitation behavior, we fit the \gls*{RIXS} spectra in the low-energy region with three peaks (a Gaussian function for the elastic peak and two \gls*{DHO} functions for the magnetic-excitation peak and higher-order magnetic-excitation peak) plus a quadratic background. The full-width at half-maximum of the Gaussian function was fixed to the instrument energy resolution of $32$~meV, determined from a reference measurement on a multilayer heterostructure sample with a strong elastic-scattering signal. The \gls*{DHO} function used to model the RIXS intensity $S(\Q{},\omega)$ as a function of $\Q{}$ and energy $\omega$ is
\begin{equation}
    S(\Q{},\omega)=\frac{\omega \chi_Q}{1-\exp(-\omega/k_B T)}\cdot \frac{2z_Q f_Q}{(\omega^2-f_Q^2)^2+(\omega z_Q)^2}
\end{equation}
where $f_Q$ is the undamped energy, $\chi_Q$ is the oscillator strength, $z_Q$ is the damping factor, $k_B$ is the Boltzmann constant, and $T$ is the temperature. We convolved these \gls*{DHO} functions with a Gaussian resolution function to describe the magnetic-excitation and higher-order magnetic-excitation peaks. For the background, we used the following piecewise function:
\begin{equation}
    B(\Q{},\omega) = \begin{cases}
            b_Q & \text{if } \omega < 0 \\
            a_Q \cdot \omega^2+b_Q & \text{if } \omega \geq 0
\end{cases}
\end{equation}
which captures the elevated background arising from the broad continuum centered around $0.4$~eV. Given the broad nature of the higher-order magnetic feature, we found that its peak position and width showed negligible momentum dependence. Consequently, the associated fit parameters were fixed at their best overall values ($f_Q = 160$~meV and $z_Q = 160$~meV) for all spectra. Similarly, the magnetic-excitation peak width in the $(H0L)$ plane was fixed to its best overall value of $z_Q = 37$~meV. As shown in Figs.~\ref{fig:SI_fits_1313_HHL}--\ref{fig:SI_fits_4310_H0L}, our fitting approach provides a satisfactory description of the \gls*{RIXS} spectra.

\clearpage
\section{Modeling of the magnetic excitations}
\label{sec:sm_magnetic_excitations}

Magnetic-excitation spectra were calculated using the entangled-units formalism available in the Sunny.jl package \cite{Dahlbom2025Sunny}. This formalism enables the calculation of generalized linear spin-wave spectra for systems composed of strongly coupled clusters of spins, here taken to be pairs of $S=1$ Ni moments coupled by the dominant $J_{c}$ Heisenberg exchange. The dimension of the local Hilbert space on each of these bonds is nine (the product of two $S=1$ spins), corresponding to spin-singlet, triplet, and quintet states. The state of the system is then approximated as a product of nine-level quantum states (formally, SU(9) coherent states), each assigned to a bond. This starting point contrasts with a standard spin-wave calculation, which instead begins with a product of dipoles (formally, SU(2) coherent states), each assigned to a crystallographic site. The ground state is found by numerically minimizing the expectation value of the spin Hamiltonian in this product of bond states. The optimization yields a classical ground state in the sense of Ref.~\cite{zhang2021classical}. The local operators of the spin Hamiltonian are rotated so that the local reference state corresponds to the ground state (an SU(9) transformation) before applying a generalized Holstein--Primakoff transformation \cite{muniz2014generalized}. The resulting spin-wave Hamiltonian is finally truncated to quadratic order in the bosonic operators and para-diagonalized \cite{colpa1978diagonalization}. This procedure amounts to performing a multi-flavor linear spin-wave calculation on top of the well-established bond-operator formalism frequently used for dimerized systems \cite{matveev1974quantum, sachdev1990bond, chubukov1991spontaneous, normand2011complete, toth2012competition}, where the bosonic modes correspond precisely to triplon (or higher-order multiplet) excitations. A detailed calculation of this type is provided in Ref.~\cite{Dahlbom2024Classical}. The only generalizations required here are a square lattice populated with $S=1$ (rather than $S=1/2$) spins and the addition of longer-range Heisenberg interactions.

For bilayer LNO-2222, the magnetically active sites in the two \ce{NiO2} layers of each bilayer were treated as an entangled unit. The in-plane Hamiltonian includes exchange paths up to third-nearest neighbors, as shown in Fig.~4(a) of the main text. $J_1$ couples to a non-magnetic site and was therefore omitted. The interlayer coupling within the bilayer is denoted by $J_c$. For the trilayer model, an effective dimer model was used in which the two outer \ce{NiO2} layers are magnetically active and the middle layer is assumed to be non-magnetic. We note that although we use \ce{La4Ni3O10} as an example here, this trilayer model also applies to the LNO-1313 case since both materials exhibit essentially the same magnetic-excitation dispersions. The coupling between the two outer-layer sites is denoted by $J_{c}'$. Its bond length is approximately twice that of $J_{c}$ in LNO-2222, giving rise to the different $L^*$-dependent spectral-weight periodicities shown in Fig.~4(c) and (f) of the main text. As in the main text, we define the effective coordinate $L^*=Ld/c$, where $d\approx 3.9$~\AA{} is the separation between neighboring \ce{NiO2} layers in the structure and $c$ is the crystallographic lattice constant.

The magnetic ground states were obtained by numerical energy minimization, followed by entangled-unit generalized linear spin-wave calculations of the dynamical spin structure factor. Two stripe domains related by a $90^\circ$ in-plane rotation were calculated explicitly and combined as an incoherent 50/50 average. The calculated spectra were evaluated at the experimental momentum and energy coordinates and convolved with the corresponding instrumental resolution.

\begin{figure}
\includegraphics{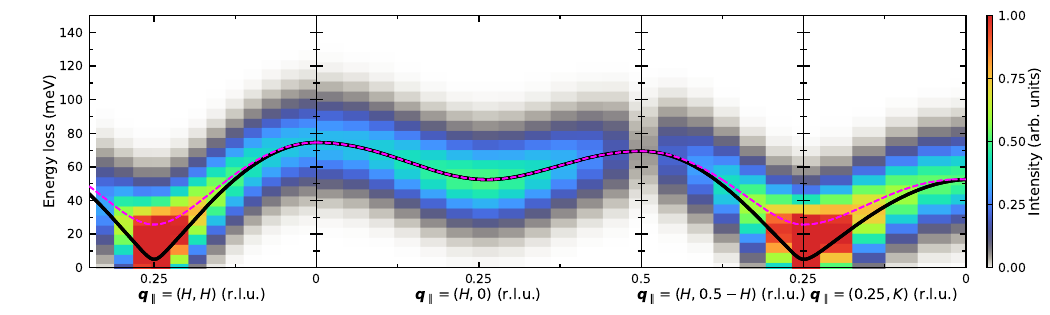}
\caption{Simulated in-plane magnetic-excitation dispersion in bilayer LNO-2222. The simulation is the best fit to the data from Ref.~\cite{Chen2024Electronic}. The intensity maps are an incoherent average of contributions from two equally populated, orthogonal magnetic stripe domains, broadened by the instrument resolution and plotted on the same $\bm{Q}$ grid as the experimental data. The curves show the visible magnetic-excitation bands from each domain: solid black lines represent the domain with the spin-ordering wavevector along $(H,H)$, and dashed magenta lines represent the other domain, which is rotated by $90^{\circ}$.
}
\label{fig:SI_additional_SW_calculations_2222}
\end{figure}

For each material, we simultaneously fitted the available in-plane and $L^*$-dependent \gls*{RIXS} spectra. For LNO-2222, the fit included four in-plane datasets and one $L^*$-dependent dataset at a fixed in-plane momentum reported in Ref.~\cite{Chen2024Electronic}. Figure~4(b) of the main text shows two of the fitted in-plane directions, while Fig.~\ref{fig:SI_additional_SW_calculations_2222} shows the complete in-plane trajectory. For \ce{La4Ni3O10}, two in-plane datasets and one $L^*$-dependent dataset were fitted. Elastic-line regions were excluded because their intensities can depend strongly on surface conditions and are not directly comparable with the calculated magnetic intensity near the spin ordering wavevector. For LNO-2222, the independently determined magnetic gaps from a recent neutron-scattering report \cite{Chen2026Nature} were included as constraints in the fitting objective. At $\mathbf{Q}_{\mathrm{SO}}=(0.25,0.25)$, the calculated gaps associated with the two stripe domains were constrained to $5$ and $25$~meV, respectively. No corresponding gap constraint was imposed for \ce{La4Ni3O10}.

Exchange parameters were optimized using Latin-hypercube sampling followed by Nelder-Mead refinement of the best candidate solutions. Parameter uncertainties were estimated using profile-loss scans, with each exchange parameter fixed successively while the remaining free parameters were re-optimized. The reported error bars correspond to approximate $95\%$ profile-loss intervals.

\clearpage
\bibliography{refs}